\documentclass[11pt, a4paper]{article}
\usepackage{graphicx,color}
\usepackage{jheppub,amsbsy,amssymb,latexsym,amsfonts,amsmath,epsfig,multicol}

\allowdisplaybreaks

\title{Anisotropic Landau-Lifshitz sigma models from $q$-deformed AdS$_5\times$S$^5$ superstrings}

\author{Takashi Kameyama,}
\author{Kentaroh Yoshida}

\affiliation{Department of Physics, Kyoto University, 
Kyoto 606-8502, Japan} 

\emailAdd{kame@gauge.scphys.kyoto-u.ac.jp}
\emailAdd{kyoshida@gauge.scphys.kyoto-u.ac.jp}

\abstract{
We consider bosonic subsectors of the $q$-deformed AdS$_5\times$S$^5$ superstring action 
and study the classical integrable structure of anisotropic Landau-Lifshitz sigma models (LLSMs)  
derived by taking fast-moving string limits. 
The subsectors are 1) deformed AdS$_3\,\times\,$S$^1$ and 2) R\,$\times$\,deformed S$^3$\,.  
The cases 1) and 2) lead to a time-like warped $SL(2)$ LLSM and a squashed S$^3$ LLSM, respectively.   
For each of them, we construct an infinite number of non-local conserved charges 
and show a quantum affine algebra at the classical level. 
Furthermore, a pp-wave like limit is applied for the case 1). 
The resulting system is a null-like warped $SL(2)$ LLSM and 
exhibits a couple of Yangians through non-local gauge transformations 
associated with Jordanian twists. 
}


\arxivnumber{1405.4467}

\begin{document}

\maketitle

\section{Introduction}

The AdS/CFT correspondence \cite{M} is a particular class of dualities between string (gravity) theories 
and gauge theories. It is firmly supported by an enormous amount of works to date and its various aspects have been 
elucidated. In particular, the discovery of an integrable structure behind it \cite{review} is 
a triumph of modern theoretical and mathematical physics.  

\medskip 

A significant feature of AdS/CFT is that the AdS$_5\times$S$^5$ background 
is represented by a supercoset 
\begin{eqnarray}
\mbox{AdS}_5\times\mbox{S}^5 = \frac{PSU(2,2|4)}{SO(1,4)\times SO(5)}\,. \label{coset}
\end{eqnarray}
It enables us to construct the Green-Schwarz string action \cite{MT} in terms of group elements 
and the string sigma model is classically integrable \cite{BPR}. The supercoset (\ref{coset}) 
enjoys the $\mathbb{Z}_4$-grading that ensures the existence of 
an infinite number of conserved charges \cite{Luscher1,BIZZ,Bernard-Yangian,MacKay}. 
Other possible cosets are classified from the consistency conditions for string backgrounds 
\cite{Zarembo-symmetric}. 

\medskip 

The next is to consider integrable deformations of AdS/CFT.  
There are two approaches to tackle this issue.  
The one is an algebraic approach based on $q$-deformations of the world-sheet S-matrix 
\cite{BK,BGM,HHM,dLRT,Arutyunov}. The deformed S-matrices are constructed in a mathematically 
consistent way. The other is a geometric approach to argue 
deformations of target spaces of string sigma models. Integrable deformations of two-dimensional 
non-linear sigma models have a long history (For classic papers, see \cite{Cherednik,FR,BFP}). 
In the present, there is a renewed interest in this subject in relation to the study of AdS/CFT.  

\medskip 

We will focus upon the latter approach here. 
Deformed target spaces are not represented by symmetric cosets, but typically 
by non-symmetric cosets (For arguments on some examples, see \cite{SYY}). 
A particularly simple and tractable example is squashed S$^3$ and its integrable structure 
has been studied intensively 
\cite{KY,KOY,KYhybrid,KY-Sch,KMY-QAA,exotic,Jordanian-KMY,ORU,BR,OU}.
In the recent, a generalization to higher dimensions has succeeded 
for arbitrary compact Lie groups and symmetric  cosets \cite{DMV}, 
by following the Yang-Baxter sigma model description \cite{Klimcik}. 

\medskip 

Just after that, a standard $q$-deformed AdS$_5\times$S$^5$ superstring action 
has been constructed with a linear R-operator of Drinfeld-Jimbo type \cite{Drinfeld1,Drinfeld2,Jimbo}  
satisfying the modified classical Yang-Baxter equation \cite{DMV-string}. 
Then the metric (in the string frame) and NS-NS two-form have been determined \cite{ABF}. 
Some special cases of the background are examined in \cite{HRT} 
and a mirror TBA is proposed in \cite{mirror}. 

\medskip 

On the other hand, there is another kind of $q$-deformations called Jordanian deformations 
\cite{R,Jordanian,KLM}. 
Jordanian deformed  AdS$_5\times$S$^5$ superstring actions have been constructed with 
linear R-operators satisfying classical Yang-Baxter equation \cite{KMY-JordanianAdSxS}.
A remarkable point is that partial deformations are possible in comparison to the standard $q$-deformation. 
In fact, as an example of deformation of AdS$_5$\,, a complete type IIB supergravity solution 
has been found in \cite{KMY-JordanianIIB} and it contains a three-dimensional Schr\"odinger spacetime 
(Sch$_3$) as a subspace. Furthermore, $\gamma$-deformed backgrounds \cite{LM,Frolov} and 
the gravity duals for non-commutative gauge theories \cite{HI,MR,DHH} have been reproduced in \cite{MY1} and 
\cite{MY2}, respectively, in the context of the Yang-Baxter sigma model. 

\medskip 

In this paper, we will concentrate on the standard $q$-deformation of the AdS$_5\times$S$^5$ superstring 
constructed in \cite{DMV-string}. A serious problem is that the metric introduced in \cite{ABF} is singular. 
In order to resolve the singularity, it would be a nice way to consider a non-relativistic limit of 
the string world-sheet. It is performed by considering  fast-moving strings \cite{Kruczenski,Rafael1,ST,SL2}. 
For simplicity, we consider two subsectors, 1) deformed AdS$_3\times$S$^1$ and 2) R$\times$ deformed S$^3$\,. 
The deformed AdS$_3$ is still singular and hence it is enough to take this subspace 
so as to argue how to avoid the singularity.

\medskip

There is another advantage of taking a non-relativistic limit from the viewpoint of the classical integrable structure.  
While non-ultra local terms appear in the current algebra in the relativistic case, those do not appear 
in the non-relativistic case. Hence infinite-dimensional symmetries generated by conserved non-local charges 
can be studied in a definite manner without ambiguities.

\medskip 

This paper is organized as follows. 
Section 2 considers fast-moving string limits of the deformed AdS$_3\times$S$^1$ and R$\times$deformed S$^3$ 
subsectors of the standard $q$-deformed AdS$_5\times$S$^5$\,.
The resulting systems are a time-like warped $SL(2)$ LLSM and a squashed S$^3$ LLSM, respectively. 
Section 3 reveals the classical integrable structure of the time-like warped $SL(2)$ LLSM.  
The Lax pair is constructed and a (classical analogue of) quantum affine algebra $U_q(\widehat{sl(2)})$ is presented.  
The classical $r$-matrix is of trigonometric type. 
In section 4 we study a pp-wave like limit of the time-like warped $SL(2)$ LLSM. 
Then the resulting system is a null-like warped $SL(2)$ LLSM.  
A direct computation leads to an exotic symmetry as in \cite{exotic} and
the classical $r$-matrix contains a deformation term. 
However, non-local gauge transformations can be performed by following \cite{Jordanian-KMY} 
so as to undo Jordanian twists. As a result, a couple of Yangians ${\mathcal Y}(sl(2))$ 
are derived and the resulting $r$-matrix is of rational. 
Section 5 is devoted to conclusion and discussion. 

\medskip 

In Appendix A, our convention of the $sl(2)$  and $su(2)$ generators are summarized. 
Appendix B explains the classical integrable structure of the squashed S$^3$ LLSM. 
Also in this case, a quantum affine algebra $U_q(\widehat{su(2)})$ is exhibited.  
The classical $r$-matrix is of  trigonometric type again.
In Appendix C, we argue the relation between boundary conditions and conserved non-local charges. 
In Appendix D, a null-like warped LLSM is derived from a time-like LLSM via a pp-wave like limit.  
In Appendix E, we give the detailed computation of non-local gauge transformations 
in undoing Jordanian twists.

\section{Fast-moving string limits of $q$-deformed AdS$_5\times$S$^5$}

In this section, we first introduce the bosonic part of the $q$-deformed AdS$_5\times$S$^5$ 
superstring action constructed in \cite{DMV-string,ABF}. 
Then it is truncated to the following subsectors\,: 
1) deformed AdS$_3\times$S$^1$ and 2) R$\times$deformed S$^3$\,.
By taking a fast-moving string limit, each of them leads to an anisotropic LLSM.

\subsection{The $q$-deformed AdS$_5\times$S$^5$ background}

The standard $q$-deformation of the AdS$_5\times$S$^5$ superstring has been constructed 
in \cite{DMV-string}. Here we are interested in the bosonic part of the deformed action,  
where the metric (in the string frame) and NS-NS two-form have been determined in \cite{ABF}. 

\medskip 

The bosonic action is composed of the metric part $S_G$ and the Wess-Zumino (WZ) term $S_{\rm WZ}$ 
that describes the coupling to an NS-NS two-form as follows: 
\begin{eqnarray}	
S &=& S_G + S_{\rm WZ}\,, \label{abf} \\ 
S_G &=& \int\!d\tau d\sigma\, \left[\,
\mathcal{L}^G_{\rm AdS} + \mathcal{L}^G_{\rm S}\,
\right]\,, \qquad 
S_{\rm WZ} = \int\!d\tau d\sigma\, \left[\,
\mathcal{L}_{\textrm{AdS}}^{\rm WZ} + \mathcal{L}_{\textrm{S}}^{\rm WZ}\,
\right]\,.  
\nonumber 
\end{eqnarray} 
Here $S_G$ and $S_{\rm WZ}$ are divided into the AdS part and the internal sphere part. 

\medskip

The metric parts $\mathcal{L}_{\textrm{AdS}}^{G}$ and $\mathcal{L}_{\textrm{S}}^{G}$ are given by, respectively,  
\begin{eqnarray}
\mathcal{L}_{\textrm{AdS}}^{G} &=& -\frac{T_{(\varkappa)}}{2}\,\eta^{\mu\nu}
\Bigl[ -\dfrac{(1+\rho^2)\,\partial_\mu t\partial_\nu t}{1-\varkappa^2\rho^2}
+\dfrac{\partial_\mu\rho\partial_\nu\rho}{(1+\rho^2)(1-\varkappa^2\rho^2)}
+\dfrac{\rho^2\,\partial_\mu\zeta\partial_\nu\zeta}{1+\varkappa^2\rho^4\sin^2\zeta}\nonumber\\ 
&&\hspace{2.0cm}+\dfrac{\rho^2\cos^2\zeta\,\partial_\mu\psi_1\partial_\nu\psi_1}{1+\varkappa^2\rho^4\sin^2\zeta}
+\rho^2\sin^2\zeta\,\partial_\mu\psi_2\partial_\mu\psi_2\Bigr]\,,\\\nonumber\\
\label{ads5}
\!\!\mathcal{L}_{\textrm{S}}^{G} &=&-\frac{T_{(\varkappa)}}{2}\,  \eta^{\mu\nu}
\Bigl[  \dfrac{(1-r^2)\,\partial_\mu\phi\partial_\nu\phi}{1+\varkappa^2 r^2}
+\dfrac{\partial_\mu r\partial_\mu r}{(1- r^2)(1+\varkappa^2r^2)}
+\dfrac{r^2\,\partial_\mu\xi\partial_\nu\xi}{1+\varkappa^2r^4\sin^2\xi}\nonumber\\ 
&&\hspace{2.0cm}+\dfrac{r^2\cos^2\xi\,\partial_\mu\phi_1\partial_\nu\phi_1}{1+\varkappa^2r^4\sin^2\xi}
+r^2\sin^2\xi\,\partial_\mu\phi_2\partial_\nu\phi_2\Bigr]\,.
\label{s5}
\end{eqnarray} 
Here the deformed AdS$_5$ part is parameterized by 
the coordinates $t\,, \psi_1\,, \psi_2\,,\zeta\,, \rho$\,. 
The deformed S$^5$ part is described by $\phi\,, \phi_1\,, \phi_2 \,, \xi\,, r$\,. 
The world-sheet metric $\eta_{\mu\nu}$ is taken as the flat metric $\eta_{\mu\nu}=(-1,+1)$ 
with the world-sheet coordinates $\sigma^{\mu}=(\sigma^0,\sigma^1)=(\tau,\sigma)$\,. 
The periodic boundary condition is imposed for the $\sigma$-direction and 
the range is taken as $\sigma \in [0,2\pi]$\,.

\medskip 

Then the deformation is measured by a real  parameter $\varkappa \in [0,\infty)$\,. 
This parameter can also be expressed in terms of another real parameter $\eta \in [0\,,1)$ like 
\begin{eqnarray}
	\varkappa=\frac{2\eta}{1-\eta^2}\,. 
\label{varkappa}
\end{eqnarray} 
When $\varkappa=0$\,, the above action is reduced to the undeformed AdS$_5\times$S$^5$ one. 
The range of the coordinates is the same as that of the ones with $\varkappa=0$\,. 
A $\varkappa$-dependent string tension $T_{(\varkappa)}$ is defined as 
\begin{eqnarray}
	T_{(\varkappa)}\equiv T(1+\varkappa^2)^{1/2}\,,\qquad T\equiv \dfrac{R^{2}}{2\pi\alpha'}\,.
\label{t_effect}
\end{eqnarray} 
Here $T$ is a dimensionless string tension with $R^2$\,, 
where $R$ is the AdS radius when $\varkappa=0$\,.

\medskip 

Finally the WZ parts are given by, respectively,  
\begin{eqnarray}
	\mathcal{L}_{\textrm{AdS}}^{WZ} &=& \frac{T_{(\varkappa)}}{2}\,\varkappa\,
\epsilon^{\mu\nu}\dfrac{\rho^4\sin2\zeta}{1+\varkappa^2\rho^4\sin^2\zeta}\,\partial_\mu\psi_1\partial_\nu\zeta\,, 
\label{wz_ads} \\  
	\mathcal{L}_{\textrm{S}}^{WZ} &=& -\frac{T_{(\varkappa)}}{2}\,\varkappa\,
\epsilon^{\mu\nu}\dfrac{r^4\sin2\xi}{1+\varkappa^2r^4\sin^2\xi}\,\partial_\mu\phi_1\partial_\nu\xi\,,
\label{wz_s}
\end{eqnarray} 
where $\epsilon^{\mu\nu}$ is the totally anti-symmetric tensor on the string world-sheet 
and it is normalized as $\epsilon^{01}=+1$\,. 
The WZ parts are proportional to $\kappa$\,, 
and hence vanish when $\varkappa=0$\,.

\subsection*{A deformed AdS$_3\times$S$^3$ subspace}

For later purpose, we consider a deformed AdS$_3\times$S$^3$ 
subspace of the $q$-deformed AdS$_5\times$S$^5$\,.
By imposing the following conditions, 
\begin{eqnarray}
\zeta=0\,,\qquad \xi=0\,, \label{cond}
\end{eqnarray}  
$S^G$ is restricted to a 
deformed AdS$_3\times$S$^3$\,.  
The metric parts 
are given by 
\begin{eqnarray}
	\mathcal{L}_{\textrm{AdS}_3}^{G} &=&-\frac{T_{(\varkappa)}}{2}\,\eta^{\mu\nu}
\Bigl[  -\dfrac{(1+\rho^2)\,\partial_\mu t\partial_\nu t}{1-\varkappa^2\rho^2}
+\dfrac{\partial_\mu\rho\partial_\nu\rho}{(1+\rho^2)(1-\varkappa^2\rho^2)}
+\rho^2\,\partial_\mu\psi_1\partial_\nu\psi_1\Bigr]\,, 
\label{ads3} \\ 
\mathcal{L}_{\textrm{S}^3}^{G} &=&-\frac{T_{(\varkappa)}}{2}\,\eta^{\mu\nu}
\Bigl[  \dfrac{(1-r^2)\,\partial_\mu\phi\partial_\nu\phi}{1+\varkappa^2 r^2}
+\dfrac{\partial_\mu r\partial_\mu r}{(1- r^2)(1+\varkappa^2r^2)}
+r^2\,\partial_\mu\phi_1\partial_\nu\phi_1\Bigr]\,.
\label{s3}
\end{eqnarray}  
Note that the $S_{\rm WZ}$ part vanishes under the condition (\ref{cond}). 

\medskip
 
To take fast-moving string limits, it is helpful to perform coordinate transformations, 
\begin{eqnarray}
	\rho\rightarrow \sinh\rho\,,\qquad r\rightarrow\sin\theta\,.
\end{eqnarray}
Then the metric parts (\ref{ads3}) and (\ref{s3}) are rewritten as
\begin{eqnarray}
	\mathcal{L}_{\textrm{AdS}_3}^{G} &=& -\frac{T_{(\varkappa)}}{2}\,\eta^{\mu\nu}
\Bigl[  -\dfrac{\cosh^2\rho\,\partial_\mu t\partial_\nu t}{1-\varkappa^2\sinh^2\rho}
+\dfrac{\partial_\mu\rho\partial_\nu\rho}{1-\varkappa^2\sinh^2\rho}
+\sinh^2\rho\,\partial_\mu\psi_1\partial_\nu\psi_1\Bigr]\,,\\
\label{ads3'}
	\mathcal{L}_{\textrm{S}^3}^{G} &=&-\frac{T_{(\varkappa)}}{2}\, \eta^{\mu\nu}
\Bigl[  \dfrac{\cos^2\theta\,\partial_\mu\phi\partial_\nu\phi}{1+\varkappa^2 \sin^2\theta}
+\dfrac{\partial_\mu \theta\partial_\mu \theta}{1+\varkappa^2\sin^2\theta}+\sin^2\theta\,\partial_\mu\phi_1\partial_\nu\phi_1\Bigr]\,.
\label{s3'}
\end{eqnarray} 
This truncated action will be the starting point of our later arguments.

\subsection{A fast-moving string limit of deformed AdS$_3\times$S$^1$}

Let us consider the string action on a deformed AdS$_3\times$S$^1$ subsector of the truncated action. 
By setting that $\theta=0$\,, the Lagrangian is given by
\begin{eqnarray}
	\mathcal{L}_{\textrm{AdS}_3\times \textrm{S}^1}^{G} &=&-\frac{T_{(\varkappa)}}{2}\, \eta^{\mu\nu}
\Bigl[  -\dfrac{\cosh^2\rho\,\partial_\mu t\partial_\nu t}{1-\varkappa^2\sinh^2\rho}
+\dfrac{\partial_\mu\rho\partial_\nu\rho}{1-\varkappa^2\sinh^2\rho}\nonumber \\ 
&& \hspace*{3.0cm}+\sinh^2\rho\,\partial_\mu\psi_1\partial_\nu\psi_1+\partial_\mu\phi\partial_\nu\phi \Bigr]\,, 
\label{ta1}
\end{eqnarray} 
where an S$^1$ circle is described by $\phi$\,.

\medskip

To derive an LLSM from this subsector, let us perform a coordinate transformation, 
\begin{eqnarray}
	\psi_1=\tilde{\psi} +t\,,\qquad \phi=\tilde{\phi}+t\,,\qquad \rho=\frac{\tilde{\rho}}{2}\,.
\end{eqnarray}  
Then the Lagrangian (\ref{ta1}) is rewritten as
\begin{eqnarray}
\mathcal{L}_{\textrm{AdS}_3\times\textrm{S}^1}^{G} &=& -\frac{T_{(\varkappa)}}{2}\,\eta^{\mu\nu}
\biggl[ -\frac{\varkappa^2\sinh^2\frac{\tilde{\rho}}{2}\cosh^2\frac{\tilde{\rho}}{2}}{1-\varkappa^2\sinh^2\frac{\tilde{\rho}}{2}}\,
\partial_\mu t\partial_\nu t+2\Bigl(\sinh^2\frac{\tilde{\rho}}{2}\partial_\mu\tilde{\psi}+ \partial_\mu\tilde{\phi}\Bigr)\partial_\nu t\nonumber \\ 
&& \hspace*{2cm}
+\frac{1}{4}\frac{\partial_\mu \tilde{\rho}\partial_\nu \tilde{\rho}}{1-\varkappa^2\sinh^2\frac{\tilde{\rho}}{2}} 
+\sinh^2\frac{\tilde{\rho}}{2}\partial_\mu \tilde{\psi}\partial_\nu \tilde{\psi}+ \partial_\mu\tilde{\phi}\partial_\nu\tilde{\phi} \biggr]\,.
\end{eqnarray}
With the static gauge 
\[
t = \kappa \tau\,,
\] 
the following fast-moving string limit is taken, 
\begin{eqnarray}
	\varkappa\rightarrow 0\,,\quad \dot{X}^\mu \rightarrow 0\,,\quad \kappa \rightarrow \infty  \quad \textrm{with} \quad \kappa\varkappa, \quad \kappa\dot{X}^\mu:~~ \textrm{fixed}\,.
\label{fast}
\end{eqnarray}
Note that the above limit (\ref{fast}) contains a further condition on $\kappa\varkappa$ as well as the standard one 
discussed in \cite{Kruczenski,ST,SL2}. 

\medskip

After all, the resulting action is given by 
\begin{eqnarray}
S &=&  \frac{T}{2}\int\!\!d\tau d\sigma\,
\Bigl[-\frac{1}{4}\kappa^2\varkappa^2\sinh^2\tilde{\rho} +\kappa\bigl[(\cosh\tilde{\rho}-1)\dot{\tilde{\psi}}+2 \dot{\tilde{\phi}}\bigr]\nonumber \\ 
&& \hspace*{2cm}
-\frac{\acute{\tilde{\rho}}^2}{4}
-\sinh^2\frac{\tilde{\rho}}{2}\,\acute{\tilde{\psi}}^2-\acute{\tilde{\phi}}^2\Bigr]\,.
\label{ads3fast}
\end{eqnarray}
Here $T_{(\varkappa)}$ has been reduced to the dimensionless tension $T$ after performing the limit (\ref{fast})\,.
A remarkable point is that the system (\ref{ads3fast}) has no singular term in comparison to the original metric.

\medskip

The Virasoro constraints are also rewritten under the limit (\ref{fast}). 
To the leading order in $\kappa$, the one of the Virasoro constraints becomes 
\begin{eqnarray}
	0 = \kappa\bigl[(\cosh\tilde{\rho}-1)\acute{\tilde{\psi}}+2 \acute{\tilde{\phi}}\,\bigr]\,.
\label{v2}
\end{eqnarray}
By eliminating $\acute{\tilde{\phi}}$ from (\ref{ads3fast}) with (\ref{v2}),  
the leading-order action is given by 
\begin{eqnarray}
S &=& \frac{T}{2}\int\!\!d\tau d\sigma\,
\Bigl[-\frac{1}{4}\kappa^2\varkappa^2\sinh^2\tilde{\rho} 
+ \kappa \bigl[(\cosh\tilde{\rho}-1)\dot{\tilde{\psi}}+2 \dot{\tilde{\phi}}\bigr] 
\nonumber \\ 
&& \hspace*{2.0cm}-\frac{1}{4}\bigl(\acute{\tilde{\rho}}^2+\sinh^2\tilde{\rho}\,\acute{\tilde{\psi}}^2\bigr)\Bigr]\,.
\label{llads3}
\end{eqnarray}
The above action is simple and contains just a single deformation term (only the first term). 

\subsubsection*{A comparison to the time-like warped AdS$_3$ case}

It is valuable to rewrite the action (\ref{llads3}) by introducing new parameters,  
\begin{eqnarray}
	C\equiv \frac{\varkappa^2}{4}\geq0\,,\qquad L \equiv \dfrac{R^{2}\kappa}{2\pi\alpha'}\,, \qquad \lambda \equiv \dfrac{R^{4}}{\alpha '^{2}}\,.
\end{eqnarray}
Then the action (\ref{llads3}) is rewritten as
\begin{eqnarray}
S &=& \dfrac{L}{2} \!\int\!\! dt d\sigma\biggl[-C\cosh^{2}{\tilde{\rho}} 
+ \bigl[(\cosh{\tilde{\rho}}-1)\partial_{t}\tilde{\psi} +2\partial_{t}\tilde{\phi}\bigr] \nonumber \\ 
&& \hspace*{2.0cm}
		- \dfrac{\lambda}{16\pi^{2}L^{2}}\bigl[(\partial_{\sigma}\tilde{\rho})^{2} 
+ \sinh^{2}{\tilde{\rho}}\,(\partial_{\sigma}\tilde{\psi})^{2}\bigr]\biggr]\,. \label{res1}
\end{eqnarray} 
Here $\tau$ is replaced by $t$ through $t=\kappa\tau$\,.  
The deformation term has also been rewritten like $-C \sinh^2\tilde{\rho} \rightarrow -C\cosh^2\tilde{\rho} +C$  
and then the constant term $C$ has been dropped off. 

\medskip 

As a result, the derived action (\ref{res1}) agrees precisely with a fast-moving string  limit 
of time-like warped AdS$_3$$\times$S$^1$ string sigma model\cite{KameYoshi}. 
Hence we call it {\it time-like warped $SL(2)$ LLSM}.

\subsection{A fast-moving string limit of R$\times$deformed S$^3$}

We next consider the string action on an R$\times$deformed S$^3$ subsector of 
the truncated action. 
By setting that $\rho=0$\,, the Lagrangian is given by
\begin{eqnarray}
	\mathcal{L}_{\textrm{R}\times\textrm{S}^3}^{G}\!\! =  -\frac{T_{(\varkappa)}}{2}\,
\eta^{\mu\nu}\!\Bigl[  -\partial_\mu t\partial_\nu t+\dfrac{\cos^2\theta\,\partial_\mu\phi\partial_\nu\phi}{1+\varkappa^2 \sin^2\theta}
+\dfrac{\partial_\mu \theta\partial_\mu \theta}{1+\varkappa^2\sin^2\theta}
+\sin^2\theta\,\partial_\mu\phi_1\partial_\nu\phi_1\Bigr]\,. 
	\label{ac}
\end{eqnarray} 
Here the time coordinate in the AdS part is included as R. 
It should be remarked that the reduced system (\ref{ac}) is not singular due to the condition $\rho=0$\,, 
even though the time direction has been included. 

\medskip

As in the previous subsection, let us first perform the coordinate transformation, 
\begin{eqnarray}
	\phi=\varphi_1 +\varphi_2\,,\qquad \phi_1=\varphi_1 -\varphi_2\,,\qquad\varphi_1=t+\tilde{\varphi_1}\,.
\end{eqnarray}  
Then the Lagrangian (\ref{ac}) is rewritten as
\begin{eqnarray}
	\mathcal{L}_{\textrm{R}\times\textrm{S}^3}^{G} &=& 
-\frac{T_{(\varkappa)}}{2}\,\eta^{\mu\nu}
\biggl[ -\frac{\varkappa^2\sin^2\theta\cos^2\theta}{1+\varkappa^2\sin^2\theta}\,
\partial_\mu t\partial_\nu t+\Bigl(1-\frac{\varkappa^2\sin^2\theta}{1+\varkappa^2\sin^2\theta}\Bigr) 
\partial_\mu \theta\partial_\nu \theta\nonumber \\ 
&& \hspace*{0.5cm}
 +2\biggl\{\Bigl(1-\frac{\varkappa^2\sin^2\theta\cos^2\theta}{1+\varkappa^2\sin^2\theta}\Bigr)\,
\partial_\mu \tilde{\varphi}_1+\Bigl(\cos2\theta
-\frac{\varkappa^2\sin^2\theta\cos^2\theta}{1+\varkappa^2\sin^2\theta}\Bigr) \,\partial_\mu \varphi_2\biggr\}
\partial_\nu t\nonumber \\ 
&& \hspace*{1.5cm}
+\Bigl(1-\frac{\varkappa^2\sin^2\theta\cos^2\theta}{1+\varkappa^2\sin^2\theta}\Bigr)( \partial_\mu\tilde{\varphi}_1\partial_\nu\tilde{\varphi}_1+\partial_\mu \varphi_2\partial_\nu \varphi_2 ) \nonumber \\ 
&& \hspace*{2.5cm}
+2\Bigl(\cos2\theta-\frac{\varkappa^2\sin^2\theta\cos^2\theta}{1+\varkappa^2\sin^2\theta}\Bigr) \partial_\mu\tilde{\varphi}_1\partial_\nu \varphi_2\biggr]\,.
\end{eqnarray}
This Lagrangian is drastically simplified with the static gauge 
\[
t = \kappa \tau
\] 
and by taking the fast-moving string limit (\ref{fast}). The resulting action is 
\begin{eqnarray}
S &=&\frac{T}{2}\int\!\!d\tau d\sigma\,
\Bigl[-\frac{1}{4}\kappa^2\varkappa^2\sin^22\theta +2\kappa\bigl[\dot{\tilde{\varphi}}_1
+\cos2\theta \dot{\varphi}_2\bigr] \nonumber \\ 
&& \hspace*{2.3cm}-\acute{\theta}^{2}- \acute{\tilde{\varphi}}_1^{2} -\acute{\varphi}_2^{2}  
-2\cos2\theta\,\acute{\tilde{\varphi}}_1\acute{\varphi}_2\Bigr]\,.
\label{s3fast}
\end{eqnarray}
Here $T_{(\varkappa)}$ is replaced by $T$ after taking the limit (\ref{fast})\,, again.

\medskip

The Virasoro constraints are also changed under the limit (\ref{fast}). 
To the leading order in $\kappa$, the one of the Virasoro constraints is rewritten as 
\begin{eqnarray}
	0 = 2\kappa[\acute{\tilde{\varphi}}_1 + \cos2\theta\acute{\varphi}_2]\,.
\label{v1}
\end{eqnarray}
By eliminating $\acute{\tilde{\varphi}}_1$ from (\ref{s3fast}) with (\ref{v1}),  
the leading-order action is given by 
\begin{eqnarray}
S =\frac{T}{2}\int\!\!d\tau d\sigma\,
\Bigl[-\frac{1}{4}\kappa^2\varkappa^2\sin^22\theta +2\kappa\bigl[\dot{\tilde{\varphi}}_1+\cos2\theta \dot{\varphi}_2\bigr]-\acute{\theta}^{2}- \sin^22\theta\acute{\varphi}_2^{2} \Bigr]\,.
\label{lls3}
\end{eqnarray}
The resulting action is very simplified again.

\subsubsection*{A comparison to the squashed S$^3$ case}

It is worth rewriting the action (\ref{lls3}) by introducing new variables, 
\begin{eqnarray}
	\varphi_2=-\frac{1}{2}\tilde{\phi}\,,\qquad \theta=\frac{1}{2}\tilde{\theta}\,,
\end{eqnarray} 
and new parameters 
\begin{eqnarray}
	C\equiv \frac{\varkappa^2}{4}\geq0\,,\qquad L \equiv \dfrac{R^{2}\kappa}{2\pi\alpha'}\,, \qquad \lambda \equiv \dfrac{R^{4}}{\alpha '^{2}}\,.
\end{eqnarray}
Then the action is rewritten as
\begin{eqnarray}
S &=& \dfrac{L}{2} \int\!\! dt d\sigma\,\biggl[C\cos^{2}{\tilde{\theta}} 
- \bigl[\cos{\tilde{\theta}}\partial_{t}\tilde{\phi} -2\partial_{t}\tilde{\varphi}_1\bigr]\nonumber \\ 
&& \hspace*{2.5cm} - \dfrac{\lambda}{16\pi^{2}L^{2}}\bigl[(\partial_{\sigma}\tilde{\theta})^{2} 
+ \sin^{2}{\tilde{\theta}}(\partial_{\sigma}\tilde{\phi})^{2}\bigr]\biggr]\,. \label{res2}
\end{eqnarray} 
Here $\tau$ is replaced by $t$ through $t=\kappa\tau$\,.  
The deformation term has been rewritten as $-C\sin^2\tilde{\theta} \rightarrow C\cos^2\tilde{\theta}-C$ 
and then the constant term $-C$ has been dropped off. 

\medskip 

Finally, the resulting action (\ref{res2}) agrees exactly with 
a fast-moving string  limit of R$\times$squashed S$^3$ string sigma model \cite{Wen}.
Hence we refer to it as  the {\it squashed S$^{\,3}$ LLSM}.

\section{Integrability of time-like warped $SL(2)$ LLSM}

Let us argue the classical integrability of the time-like warped $SL(2)$ LLSM.
The related infinite-dimensional symmetry is also discussed by explicitly constructing 
an infinite number of conserved non-local charges. 
In section 2 the periodic boundary condition has been imposed  for the 
spatial direction of the string world-sheet.
In the following sections, however, the string world-sheet is supposed to be spatially infinite 
in order to argue an infinite-dimensional symmetry based on non-local charges. 
In fact, the fast-moving string limit implies a decompactification limit of the $\sigma$-direction 
(For example, see the argument in 2.2 of \cite{AF})\,.  
For the case of the squashed S$^3$ LLSM, see Appendix B.

\subsection{The classical action and Lax pair}

The classical action of the time-like warped $SL(2)$ LLSM is given by 
\begin{eqnarray}
S&=&\frac{L}{2}\int^\infty_{-\infty}\!\!\!dtdx\,\biggl[-C\cosh^2\rho+\left(\cosh\rho-1\right)\partial_t\psi  \nonumber \\
&& \hspace*{3cm}
-\frac{\lambda}{16\pi^2L^2}\left[\left(\partial_x\rho\right)^2+\sinh^2\rho\left(\partial_x\psi\right)^2\right]\biggr]\,. 
\end{eqnarray}
The world-sheet is a (1+1)-dimensional spacetime spanned by $t$ and $x$ and the spatial direction is infinite. 
The system is non-relativistic because the action contains the first order in time derivative 
and the second order in spatial derivative. 
The deformation parameter $C$ is restricted to $C\geq 0$\,.  
When $C=0$\,, an isotropic $SL(2)$ LLSM is reproduced. 

\medskip 

It is convenient to introduce a vector representation $n^a$\,, 
\begin{eqnarray}
n^0=-\cosh\rho\,, \qquad
n^1=\sinh\rho\sin\psi\,, \qquad
n^2=\sinh\rho\cos\psi\,,
\end{eqnarray}
satisfying the following relation,
\begin{eqnarray}
\left(n^0\right)^2-\left(n^1\right)^2-\left(n^2\right)^2=1\,.
\end{eqnarray}
Then the classical equations of motion are rewritten as 
\begin{eqnarray}
&&\partial_t n^0=\frac{\lambda}{8\pi^2L^2}\left(n^1\partial_x^2 n^2-n^2\partial_x^2 n^1\right)\,, 
\nonumber \\
&&\partial_t n^1=\frac{\lambda}{8\pi^2L^2}\left(n^0\partial_x^2 n^2-n^2\partial_x^2 n^0\right)-2Cn^0n^2\,, 
\nonumber \\
&&\partial_t n^2=\frac{\lambda}{8\pi^2L^2}\left(n^1\partial_x^2 n^0-n^0\partial_x^2 n^1\right)+2Cn^0n^1\,.
\label{eomXXZ}
\end{eqnarray}
These are identical to the Landau-Lifshitz equations
\begin{eqnarray}
	\partial_{t} n_a =  \varepsilon_{abc} n^b \left( 
\dfrac{\lambda}{8\pi^{2}L^{2}}\partial_{x}^{2} n^c + \mathcal{J}^{c}_{~d} n^d\right)\qquad (a=0,1,2)\,,
\end{eqnarray}
with an anisotropic matrix $\mathcal{J}$ 
\begin{eqnarray}
\mathcal{J}^{a}_{~b} =  \textrm{diag}( j +2C, j, j) \qquad (j :~\textrm{an~arbitrary~const.})\,.
\end{eqnarray}
Here we have introduced the totally anti-symmetric tensor $\varepsilon_{abc}$ 
with $\varepsilon_{012} = - 1$. The $sl(2)$ indices are raised and lowered 
with $\gamma^{ab} = \textrm{diag}(-1\,, +1\,, +1)$ and its inverse, respectively. 

\paragraph{Lax pair.} 
The Lax pair of the time-like warped  $SL(2)$ LLSM  is represented by \cite{FT}
\begin{eqnarray}
U(t,x;z)&=&\dfrac{i\alpha}{\sinh z} \left[-\cosh z\, n^0T^0 + n^1T^1+ n^2T^2 \right] \,, \\
V(t,x;z)&=&\dfrac{i\beta}{\sinh z} \left[-\cosh z\left(n^1\partial_x n^2-n^2\partial_x n^1\right)T^0 +\left(n^0\partial_x n^2-n^2\partial_x n^0\right) T^1 \right. \nonumber \\
&&\hspace{1.5cm}\left.+\left(n^1\partial_x n^0-n^0\partial_x n^1\right)T^2 \right] \nonumber \\
&&+\dfrac{\alpha\beta}{\sinh^2 z} \left[- n^0T^0 + \cosh z\,n^1T^1+\cosh z\,n^2T^2 \right]\,, \notag 
\end{eqnarray}
with a spectral parameter $z \in \mathbb{C}$ and new parameters,
\begin{eqnarray}
\alpha \equiv \dfrac{4\pi L}{\sqrt{\lambda}}\sqrt{C}\,, \qquad 
\beta \equiv \dfrac{\sqrt{\lambda}}{2\pi L} \sqrt{C}\,.
\end{eqnarray}
The equations of motion (\ref{eomXXZ}) are reproduced from the commutation relation,  
\begin{eqnarray}
\Bigl[\partial_t-V(t,x;z),\partial_x-U(t,x;z)\Bigr]=0\,. 
\label{flatness_Lax_XXZ}
\end{eqnarray}

\paragraph{Monodromy matrix.} The monodromy matrix is defined as 
\begin{eqnarray}
M(z) \equiv {\rm P}\exp\left[\int^\infty_{-\infty}\!\!\!dx~U(t,x;z)\right]\,, 
\end{eqnarray}
where P denotes the path ordering. 
Due to the condition (\ref{flatness_Lax_XXZ})\,, $M(z)$ is a conserved quantity,
\begin{eqnarray}
\frac{d}{dt}M(z)=0\,. 
\end{eqnarray}
Thus the expansion of $M(z)$ in terms of $z$ leads to an infinite number of conserved charges. 
The resulting algebra of the charges depends on the expansion point. 
For example, the expansions around $z=\pm\infty$ leads to a quantum affine 
algebra $U_q(\widehat{sl(2)})$\,. 
We will elaborate a classical realization of it
in the next subsection. 

\subsection{The standard $q$-deformation of $sl(2)$}

We will show that a $q$-deformed $sl(2)$ is realized in the time-like warped $SL(2)$ LLSM. 

\medskip 

While the $SL(2)$ symmetry is realized in an isotropic $SL(2)$ LLSM,  
it is broken to $U(1)$ due to the non-vanishing $C$\,. 
The remaining $U(1)$ charge is given by  
\begin{eqnarray}
Q^0=-\dfrac{L}{2}\int^\infty_{-\infty}\!\!\!dx~n^0(x)\,.  
\end{eqnarray}

\medskip 

Here it is worth noting that the broken components of $SL(2)$\,,
$1$ and $2$ are still realized as non-local 
symmetries even when $C\neq 0$\,, 
as in the original system before taking the fast-moving string  limit.

\medskip 

In order to show this fact, it is convenient to introduce $n^{\widehat{\pm}}$ defined as 
\begin{eqnarray}
n^{\widehat{\pm}} \equiv \frac{n^1 \pm i n^2}{\sqrt{2}}\,. 
\end{eqnarray}
For the conservation of non-local charges, boundary conditions are sensitive.  
We take a rapidly damping condition so that $n^{\widehat{\pm}} $ vanish at the spatial infinities. 
This condition enables us to construct conserved non-local charges, as shown in Appendix C.

\medskip

The conserved non-local charges are given by 
\begin{eqnarray}
Q^{\widehat{\pm}}=-\dfrac{L}{2}\int^\infty_{-\infty}\!\!\!dx~{\rm e}^{\alpha\chi(x)}n^{\widehat{\pm}}(x)\,, 
\end{eqnarray}
where $\chi(x)$ is a non-local field defined as 
\begin{eqnarray}
\chi(x) \equiv \frac{1}{2}\int^\infty_{-\infty}\!\!\!dy~\epsilon(x-y)n^0(y)\,,
\end{eqnarray}
and $\epsilon(x)$ is the signature function defined as
\begin{eqnarray}
\epsilon(x)\equiv\theta(x)-\theta(-x) 
\end{eqnarray}
with the step function $\theta(x)$\,. It is helpful to introduce the following relations: 
\begin{eqnarray}
\partial_x\chi=n^0\,, \qquad
\partial_t\chi=i\frac{\lambda}{8\pi^2L^2}\left(n^{\widehat{+}}\partial_x n^{\widehat{-}}-n^{\widehat{-}}\partial_x n^{\widehat{+}}\right)\,. 
\end{eqnarray}

\medskip

Then the next is to compute the Poisson brackets of $Q^0$ and $Q^{\widehat{\pm}}$\,. 
The Poisson brackets for $n^a\,(a=0,\widehat{\pm})$ are given by 
\begin{eqnarray}
\Big\{n^a(x),n^b(y)\Bigr\}_{\rm P}=-i\frac{2}{L}\,\varepsilon^{ab}_{~~c}n^c(x)\delta(x-y)\,. 
\label{Poisson_n}
\end{eqnarray}
Note that non-ultra local terms are not contained in comparison to the current algebra of 
principal chiral models. 

\medskip 

With the brackets (\ref{Poisson_n})\,, 
the brackets of $Q^0$ and $Q^{\widehat{\pm}}$ can be evaluated as
\begin{eqnarray}
&&\Bigl\{Q^{\widehat{\pm}},Q^0\Bigr\}_{\rm P}=\pm iQ^{\widehat{\pm}}\,,\\
&&\Bigl\{Q^{\widehat{+}},Q^{\widehat{-}}\Bigr\}_{\rm P}= i\dfrac{L}{2\alpha} \sinh\Bigl(\dfrac{2\alpha}{L}Q^0\Bigr)\,.\nonumber
\end{eqnarray}
This is a classical analogue of the standard $q$-deformation of $sl(2)$ \cite{Drinfeld2,Jimbo}. 

\medskip 

In addition, there exists another set of non-local conserved charges, 
\begin{eqnarray}
\widetilde{Q}^{\widehat{\pm}}=-\dfrac{L}{2}\int^\infty_{-\infty}\!\!\!dx~{\rm e}^{-\alpha\chi(x)}n^{\widehat{\pm}}(x)\,. 
\end{eqnarray}
These can be obtained by changing the sign of $\alpha$ in $Q^{\widehat{\pm}}$ 
and hence $Q^0$ and $\widetilde{Q}^{\widehat{\pm}}$ also generate another $q$-deformed $sl(2)$ algebra,  
\begin{eqnarray}
&&\Bigl\{\widetilde{Q}^{\widehat{\pm}},{Q}^0\Bigr\}_{\rm P}=\pm i\widetilde{Q}^{\widehat{\pm}}\,,\\
&&\Bigl\{\widetilde{Q}^{\widehat{+}},\widetilde{Q}^{\widehat{-}}\Bigr\}_{\rm P}=i \dfrac{L}{2\alpha} \sinh\Bigl(\dfrac{2\alpha}{L}Q^0\Bigr)\,.\nonumber
\end{eqnarray}
In subsection \ref{QAA:sec}, we will show that a (classical analogue of) quantum affine algebra is generated 
by $Q^0$\,, $Q^{\widehat{\pm}}$ and $\widetilde{Q}^{\widehat{\pm}}$ 
in the sense of Drinfeld's first realization \cite{Drinfeld1}.

\subsection{Monodromy expansions and higher non-local charges \label{expansion:sec}} 

An infinite number of conserved charges are obtained by expanding the monodromy matrix $M(z)$ with respect to 
a complex parameter $u = \textrm{e}^{-z}$\,. 
Here we derive the conserved charges discussed in the previous subsection 
by expanding $M(z)$\,. According to the expansion, other higher non-local charges are also obtained. 

\medskip 

Depending on the values of $u$\,, the following two expansions are possible: 
\begin{eqnarray}
&&\textrm{i}) \qquad M(z) = \textrm{e}^{\bar{q}_0} 
\exp \left[\, \sum_{n=1}^\infty u^n \bar{q}_n\, \right]  \qquad \textrm{for}\quad |u| \ll 1\,, \nonumber \\
&&\textrm{ii}) \qquad M(z) = \textrm{e}^{q_0} \exp \left[\, \sum_{n=1}^\infty u^{-n} q_n \,\right]
\qquad \textrm{for}\quad |u| \gg 1\,. \nonumber
\end{eqnarray}
Let us consider each of the two expansions below. 

\subsubsection*{Expansion i)}

The monodromy matrix is expanded like 
\begin{eqnarray}
	&&M(z) ={\rm P}\exp\Bigl[\int^\infty_{-\infty}\!\!\!dx~U(t,x;z)\Bigr]\notag \\
&&=\textrm{e}^{i\frac{2\alpha}{L}Q_{(0)}^0T^0} \textrm{P} \exp \biggl[ 2i\alpha  \int^{\infty}_{-\infty}\!\!\!dx~\Bigl\{(u+u^3)\Bigl(T^{\widehat{+}}\textrm{e}^{-\frac{\alpha}{L}Q_{(0)}^0} \textrm{e}^{\alpha \chi}n^{\widehat{-}}(x) +\,T^{\widehat{-}}\textrm{e}^{\frac{\alpha}{L}Q_{(0)}^0}\textrm{e}^{-\alpha \chi }n^{\widehat{+}}(x) \Bigr)\notag \\ &&\qquad\qquad\qquad\qquad\qquad\qquad- u^2 T^0 n^0 (x)+ \mathcal{O}(u^4) \Bigr\} \biggr]\notag \\ 
	&&= \textrm{e}^{\bar{q}_0} \biggl[ 1+ u\bar{q}_1 + u^2 \Bigl(\bar{q}_2 + \dfrac{1}{2} (\bar{q}_1)^2 \Bigr) + u^3 \Bigl(\bar{q}_3 + \dfrac{1}{2} (\bar{q}_2\bar{q}_1+ \bar{q}_1\bar{q}_2) + \dfrac{1}{6}(\bar{q}_1)^3 \Bigr) + \mathcal{O}(u^4) \biggr]\,.\notag
\end{eqnarray}
where $\bar{q}_i ~(i = 0, 1, 2, 3,...)$ are defined as 
\begin{eqnarray}
&&\bar{q}_0 \equiv i\dfrac{2\alpha}{L}T^0 Q_{(0)}^0\,, \qquad 
\bar{q}_1 \equiv -2i\dfrac{2\alpha}{L}\Bigl(T^{\widehat{+}} \textrm{e}^{-\frac{\alpha}{L}Q_{(0)}^0}Q_{(1)}^{\widehat{-}} + T^{\widehat{-}} \textrm{e}^{\frac{\alpha}{L}Q_{(0)}^0}\widetilde{Q}_{(1)}^{\widehat{+}} \Bigr)\,,\notag \\
&&\bar{q}_2 \equiv -2i\Bigl(\dfrac{2\alpha}{L}\Bigr)^2T^0\bar{Q}_{(2)}^0 \,,\qquad 
\bar{q}_3 \equiv 2i\Bigl(\dfrac{2\alpha}{L}\Bigr)^3\Bigl(T^{\widehat{+}} \textrm{e}^{-\frac{\alpha}{L}Q_{(0)}^0}Q_{(3)}^{\widehat{-}} + T^{\widehat{-}} \textrm{e}^{\frac{\alpha}{L}Q_{(0)}^0}\widetilde{Q}_{(3)}^{\widehat{+}} \Bigr)\,, \notag \\ 
&& \hspace*{3cm} \vdots 
\end{eqnarray}
Then the conserved charges are given by
\begin{eqnarray}
&&Q_{(0)}^0 = -\dfrac{L}{2}\int^\infty_{-\infty}\!\!\!dx~n^0(x)\,, \notag\\
&&Q_{(1)}^{\widehat{-}}  = -\dfrac{L}{2}\int^\infty_{-\infty}\!\!\!dx~{\rm e}^{\alpha\chi}n^{\widehat{-}}(x)\,,\quad \widetilde{Q}_{(1)}^{\widehat{+}}  = -\dfrac{L}{2}\int^\infty_{-\infty}\!\!\!dx~{\rm e}^{-\alpha\chi}n^{\widehat{+}}(x)\,,\notag \\
&&\bar{Q}_{(2)}^0 = \Bigl(-\dfrac{L}{2}\Bigr)^2 \int^\infty_{-\infty}\!\!\!dx\int^\infty_{-\infty}\!\!\!dy~\epsilon(x-y) {\rm e}^{\alpha\chi}n^{\widehat{-}}(x){\rm e}^{-\alpha\chi}n^{\widehat{+}}(y) - \dfrac{L}{2\alpha}Q_{(0)}^0\,,\notag \\
&&Q_{(3)}^{\widehat{-}} = \Bigl(-\dfrac{L}{2}\Bigr)^3 \int^\infty_{-\infty}\!\!\!dx\int^\infty_{-\infty}\!\!\!dy\int^\infty_{-\infty}\!\!\!dz~\dfrac{1}{2}\epsilon(x-y) \epsilon(x-z) {\rm e}^{-\alpha\chi}n^{\widehat{+}}(x) {\rm e}^{\alpha\chi}n^{\widehat{-}}(y) {\rm e}^{\alpha\chi}n^{\widehat{-}}(z) \notag \\&&\quad\quad\quad - \Bigl(-\dfrac{L}{2}\Bigr)^2 \int^\infty_{-\infty}\!\!\!dx\int^\infty_{-\infty}\!\!\!dy~\dfrac{L}{2\alpha}\epsilon(x-y) {\rm e}^{\alpha\chi}n^{\widehat{-}}(x)n^{0}(y)- \dfrac{1}{6}\widetilde{Q}_{(1)}^{\widehat{+}} (Q_{(1)}^{\widehat{-}})^2+ \Bigl(\dfrac{L}{2\alpha}\Bigr)^2 Q_{(1)}^{\widehat{-}}\,, \notag \\
&&\widetilde{Q}_{(3)}^{\widehat{+}} = \Bigl(-\dfrac{L}{2}\Bigr)^3 \int^\infty_{-\infty}\!\!\!dx\int^\infty_{-\infty}\!\!\!dy\int^\infty_{-\infty}\!\!\!dz~\dfrac{1}{2}\epsilon(x-y) \epsilon(x-z) {\rm e}^{\alpha\chi}n^{\widehat{-}}(x) {\rm e}^{-\alpha\chi}n^{\widehat{+}}(y) {\rm e}^{-\alpha\chi}n^{\widehat{+}}(z)\notag \\&&\quad\quad\quad +\Bigl(-\dfrac{L}{2}\Bigr)^2 \int^\infty_{-\infty}\!\!\!dx\int^\infty_{-\infty}\!\!\!dy~\dfrac{L}{2\alpha}\epsilon(x-y) {\rm e}^{-\alpha\chi}n^{\widehat{+}}(x)n^{0}(y)- \dfrac{1}{6}Q_{(1)}^{\widehat{-}} (\widetilde{Q}_{(1)}^{\widehat{+}})^2+ \Bigl(\dfrac{L}{2\alpha}\Bigr)^2\widetilde{Q}_{(1)}^{\widehat{+}}\,, \notag \\
&& \hspace*{5cm} \vdots 
\end{eqnarray}

\subsubsection*{Expansion ii)}

The next is to consider the expansion of the monodromy matrix in the region ii). 
For this purpose, it is convenient to introduce a new parameter $\tilde{u}=1/u$\,.  
Then the monodromy matrix is expanded in terms of $\tilde{u}$ like 
\begin{eqnarray}
	&&M(z) ={\rm P}\exp\Bigl[\int^\infty_{-\infty}\!\!\!dx~U(t,x;z)\Bigr]\notag \\
&&=\textrm{e}^{-i\frac{2\alpha}{L}Q_{(0)}^0T^0} \textrm{P} \exp \biggl[ 2i\alpha  \int^{\infty}_{-\infty}\!\!\!dx~\Bigl\{-(\tilde{u}+\tilde{u}^3)\Bigl(T^{\widehat{+}}\textrm{e}^{\frac{\alpha}{L}Q_{(0)}^0} \textrm{e}^{-\alpha \chi}n^{\widehat{-}}(x) + T^{\widehat{-}}\textrm{e}^{-\frac{\alpha}{L}Q_{(0)}^0}\textrm{e}^{\alpha \chi }n^{\widehat{+}}(x) \Bigr)\notag \\
&&\qquad\qquad\qquad\qquad\qquad\qquad+ \tilde{u}^2 T^0 n^0 (x)+ \mathcal{O}(\tilde{u}^4) \Bigr\} \biggr]\notag \\ 
	&&= \textrm{e}^{q_0} \biggl[ 1+ \tilde{u}q_1 + \tilde{u}^2 \Bigl(q_2 + \dfrac{1}{2} (q_1)^2 \Bigr) + \tilde{u}^3 \Bigl(q_3 + \dfrac{1}{2} (q_2q_1+ q_1q_2) + \dfrac{1}{6}(q_1)^3 \Bigr) + \mathcal{O}(\tilde{u}^4) \biggr]\,, \notag
\end{eqnarray}
where $q_i ~(i = 0, 1, 2, 3,...)$ are defined as 
\begin{eqnarray}
&&q_0 \equiv -i \dfrac{2\alpha}{L}T^0 Q_{(0)}^0\,, \qquad 
q_1 \equiv 2i\dfrac{2\alpha}{L}\Bigl(T^{\widehat{+}} \textrm{e}^{\frac{\alpha}{L}Q_{(0)}^0}\widetilde{Q}_{(1)}^{\widehat{-}}
+ T^{\widehat{-}} \textrm{e}^{-\frac{\alpha}{L}Q_{(0)}^0}Q_{(1)}^{\widehat{+}}\Bigr)\,,\notag \\
&&q_2 \equiv 2i\Bigl(\dfrac{2\alpha}{L}\Bigr)^2T^0 Q_{(2)}^0 \,,\qquad 
q_3 \equiv -2i\Bigl(\dfrac{2\alpha}{L}\Bigr)^3\Bigl(T^{\widehat{+}}\textrm{e}^{\frac{\alpha}{L}Q_{(0)}^0}\widetilde{Q}_{(3)}^{\widehat{-}}
+ T^{\widehat{-}} \textrm{e}^{-\frac{\alpha}{L}Q_{(0)}^0}Q_{(3)}^{\widehat{+}}\Bigr)\,, \notag \\ 
&& \hspace*{3cm } \vdots  
\end{eqnarray}
The conserved charges are given by
\begin{eqnarray}
&&Q_{(0)}^0 = -\dfrac{L}{2}\int^\infty_{-\infty}\!\!\!dx~n^0(x)\,, \notag\\
&&\widetilde{Q}_{(1)}^{\widehat{-}}  = -\dfrac{L}{2}\int^\infty_{-\infty}\!\!\!dx~{\rm e}^{-\alpha\chi}n^{\widehat{-}}(x)\,,\quad Q_{(1)}^{\widehat{+}}  
= -\dfrac{L}{2}\int^\infty_{-\infty}\!\!\!dx~{\rm e}^{\alpha\chi}n^{\widehat{+}}(x)\,,\notag \\
&&Q_{(2)}^0= \Bigl(-\dfrac{L}{2}\Bigr)^2 \int^\infty_{-\infty}\!\!\!dx\int^\infty_{-\infty}\!\!\!dy~\epsilon(x-y) 
{\rm e}^{\alpha\chi}n^{\widehat{+}}(x){\rm e}^{-\alpha\chi}n^{\widehat{-}}(y) - \dfrac{L}{2\alpha}Q_{(0)}^0\,,\notag \\
&&\widetilde{Q}_{(3)}^{\widehat{-}} = \Bigl(-\dfrac{L}{2}\Bigr)^3 \int^\infty_{-\infty}\!\!\!dx\int^\infty_{-\infty}\!\!\!dy\int^\infty_{-\infty}\!\!\!dz~
\dfrac{1}{2}\epsilon(x-y) \epsilon(x-z){\rm e}^{\alpha\chi}n^{\widehat{+}}(x) {\rm e}^{-\alpha\chi}n^{\widehat{-}}(y) 
{\rm e}^{-\alpha\chi}n^{\widehat{-}}(z)\notag \\&&\quad\quad\quad 
+ \Bigl(-\dfrac{L}{2}\Bigr)^2 \int^\infty_{-\infty}\!\!\!dx\int^\infty_{-\infty}\!\!\!dy~\dfrac{L}{2\alpha}
\epsilon(x-y) {\rm e}^{-\alpha\chi}n^{\widehat{-}}(x)n^{0}(y)- \dfrac{1}{6}Q_{(1)}^{\widehat{+}} (\widetilde{Q}_{(1)}^{\widehat{-}})^2 
+ \Bigl(\dfrac{L}{2\alpha}\Bigr)^2 \widetilde{Q}_{(1)}^{\widehat{-}}\,, \notag \\
&&Q_{(3)}^{\widehat{+}}= \Bigl(-\dfrac{L}{2}\Bigr)^3 \int^\infty_{-\infty}\!\!\!dx\int^\infty_{-\infty}\!\!\!dy\int^\infty_{-\infty}\!\!\!dz~
\dfrac{1}{2}\epsilon(x-y) \epsilon(x-z){\rm e}^{-\alpha\chi}n^{\widehat{-}}(x) {\rm e}^{\alpha\chi}n^{\widehat{+}}(y) 
{\rm e}^{\alpha\chi}n^{\widehat{+}}(z)\notag \\&&\quad\quad\quad -\Bigl(-\dfrac{L}{2}\Bigr)^2 
\int^\infty_{-\infty}\!\!\!dx\int^\infty_{-\infty}\!\!\!dy~\dfrac{L}{2\alpha}\epsilon(x-y) 
{\rm e}^{\alpha\chi}n^{\widehat{+}}(x)n^0(y)- \dfrac{1}{6}\widetilde{Q}_{(1)}^{\widehat{-}} (Q_{(1)}^{\widehat{+}})^2 
+ \Bigl(\dfrac{L}{2\alpha}\Bigr)^2Q_{(1)}^{\widehat{+}}\,, \notag \\
&& \hspace*{5cm} \vdots 
\end{eqnarray}
In total, we have derived all of the conserved charges presented in the previous subsection,  
in addition to higher non-local charges. 

\medskip 

It is a turn to clarify the algebraic structure of the conserved charges.
Some examples of the Poisson brackets are listed below: 
\begin{eqnarray}
&&\Bigl\{ Q_{(1)}^{\widehat{\pm}},Q_{(1)}^{\widehat{\mp}}\Bigr\}_{\rm P}= \pm i\dfrac{L}{2\alpha} \sinh\Bigl(\dfrac{2\alpha}{L}Q_{(0)}^0\Bigr)\,,\quad
\Bigl\{Q_{(1)}^{\widehat{\pm}},Q_{(1)}^{\widehat{\pm}}\Bigr\}_{\rm P}= 0\,, \quad
\Bigl\{Q_{(1)}^{\widehat{\pm}},Q_{(0)}^0\Bigr\}_{\rm P}= \pm iQ_{(1)}^{\widehat{\pm}}\,,\notag\\
&&\Bigl\{ \widetilde{Q}_{(1)}^{\widehat{\pm}},\widetilde{Q}_{(1)}^{\widehat{\mp}}\Bigr\}_{\rm P}=\pm i\dfrac{L}{2\alpha} \sinh\Bigl(\dfrac{2\alpha}{L}Q_{(0)}^0\Bigr)\,, \quad
\Bigl\{\widetilde{Q}_{(1)}^{\widehat{\pm}},\widetilde{Q}_{(1)}^{\widehat{\pm}}\Bigr\}_{\rm P}= 0\,, \quad
\Bigl\{\widetilde{Q}_{(1)}^{\widehat{\pm}},Q_{(0)}^0\Bigr\}_{\rm P}= \pm i\widetilde{Q}_{(1)}^{\widehat{\pm}}\,,\notag\\
&&\Bigl\{Q_{(1)}^{\widehat{\pm}},\widetilde{Q}_{(1)}^{\widehat{\pm}}\Bigr\}_{\rm P}= 0\,, \qquad
\Bigl\{ Q_{(1)}^{\widehat{+}},\widetilde{Q}_{(1)}^{\widehat{-}}\Bigr\}_{\rm P}= -i\dfrac{2\alpha}{L} Q_{(2)}^0\,,\qquad
\Bigl\{ Q_{(1)}^{\widehat{-}},\widetilde{Q}_{(1)}^{\widehat{+}}\Bigr\}_{\rm P}=i \dfrac{2\alpha}{L} \bar{Q}_{(2)}^0\,,\notag\\
&&\Bigl\{Q_{(2)}^{0},Q_{(1)}^{\widehat{+}}\Bigr\}_{\rm P}=i\dfrac{2\alpha}{L}\Bigl[Q_{(3)}^{\widehat{+}} +\dfrac{2}{3}\widetilde{Q}_{(1)}^{\widehat{-}}(Q_{(1)}^{\widehat{+}})^2\Bigr]\,, \notag\\
&&
\Bigl\{Q_{(2)}^{0},\widetilde{Q}_{(1)}^{\widehat{-}}\Bigr\}_{\rm P}=-i\dfrac{2\alpha}{L}\Bigl[\widetilde{Q}_{(3)}^{\widehat{-}}+\dfrac{2}{3}Q_{(1)}^{\widehat{+}}(\widetilde{Q}_{(1)}^{\widehat{-}})^2\Bigr]\,,\notag\\
&&\Bigl\{\bar{Q}_{(2)}^{0},Q_{(1)}^{\widehat{-}}\Bigr\}_{\rm P}=-i\dfrac{2\alpha}{L}\Bigl[Q_{(3)}^{\widehat{-}} +\dfrac{2}{3}\widetilde{Q}_{(1)}^{\widehat{+}}(Q_{(1)}^{\widehat{-}})^2\Bigr]\,, \notag\\
&&
\Bigl\{\bar{Q}_{(2)}^{0},\widetilde{Q}_{(1)}^{\widehat{+}}\Bigr\}_{\rm P}= i\dfrac{2\alpha}{L}\Bigl[\widetilde{Q}_{(3)}^{\widehat{+}}+\dfrac{2}{3}Q_{(1)}^{\widehat{-}}(\widetilde{Q}_{(1)}^{\widehat{+}})^2\Bigr]\,,\notag\\
&&\Bigl\{Q_{(3)}^{\widehat{+}},{Q}_{(1)}^{\widehat{+}}\Bigr\}_{\rm P}=\dfrac{i}{3}  \dfrac{2\alpha}{L}Q_{(2)}^0(Q_{(1)}^{\widehat{+}})^2\,, \quad\quad\quad
\Bigl\{\widetilde{Q}_{(3)}^{\widehat{-}},\widetilde{Q}_{(1)}^{\widehat{-}}\Bigr\}_{\rm P}=-\dfrac{i}{3}  \dfrac{2\alpha}{L}Q_{(2)}^0(\widetilde{Q}_{(1)}^{\widehat{-}})^2\,,
\notag\\
&&\Bigl\{Q_{(3)}^{\widehat{-}},{Q}_{(1)}^{\widehat{-}}\Bigr\}_{\rm P}= -\dfrac{i}{3}  \dfrac{2\alpha}{L}\bar{Q}_{(2)}^0(Q_{(1)}^{\widehat{-}})^2\,,\qquad
\Bigl\{\widetilde{Q}_{(3)}^{\widehat{+}},\widetilde{Q}_{(1)}^{\widehat{+}}\Bigr\}_{\rm P}=\dfrac{i}{3}  \dfrac{2\alpha}{L}\bar{Q}_{(2)}^0(\widetilde{Q}_{(1)}^{\widehat{+}})^2\,. \notag 
\end{eqnarray}

\subsection{Quantum affine algebra  \label{QAA:sec}}

Let us see the relation to a classical analogue of Drinfeld's first realization of quantum affine algebra \cite{Drinfeld1}. 
It is convenient to rescale the charges $Q_{(0)}^0\,,Q_{(1)}^{\widehat{\pm}}$ and $\widetilde{Q}_{(1)}^{\widehat{\pm}}$ as follows:
\begin{eqnarray}
&&H_1 \equiv 2Q_{(0)}^0\,, \qquad\qquad\qquad\quad~   H_0 \equiv -2Q_{(0)}^0\,, \notag\\
&&E_1 \equiv \Bigl(\frac{-2\alpha/L}{\sinh(\alpha/L)}\Bigr)^{1/2}Q_{(1)}^{\widehat{+}}\,, \qquad E_0 \equiv \Bigl(\frac{-2\alpha/L}{\sinh(\alpha/L)}\Bigr)^{1/2}\widetilde{Q}_{(1)}^{\widehat{-}}\,, \notag\\
&&F_1 \equiv \Bigl(\frac{-2\alpha/L}{\sinh(\alpha/L)}\Bigr)^{1/2}Q_{(1)}^{\widehat{-}}\,, \qquad~ F_0\equiv \Bigl(\frac{-2\alpha/L}{\sinh(\alpha/L)}\Bigr)^{1/2}\widetilde{Q}_{(1)}^{\widehat{+}}\,.
\end{eqnarray}
Then the Poisson brackets are evaluated as 
\begin{eqnarray}
&&i\{ H_i,H_j\}_{\rm P}= 0 \qquad (i,j=1,0)\,,\notag \\
&&i\{H_i,E_j\}_{\rm P}= A_{ij}E_j\,,\qquad i\{H_i,F_j\}_{\rm P}= -A_{ij}F_j\,,\notag \\
&&i\{E_i,F_j\}_{\rm P}= \delta_{ij} \frac{q^{H_i}-q^{-H_i}}{q-q^{-1}}\,,
\end{eqnarray}
where the generalized Cartan matrix $A_{ij}$ is given by
\begin{eqnarray}
A_{ij} =(\alpha_i,\alpha_j)= \begin{pmatrix}
			2 &-2\\
			-2 & 2
			\end{pmatrix} \quad \textrm{with} \quad 
\alpha_1= \begin{pmatrix}
			1\\
			-1
\end{pmatrix}\,,\quad
\alpha_0= \begin{pmatrix}
			-1\\
			1
\end{pmatrix}
\end{eqnarray}
and a $q$-deformation parameter is defined as
\begin{eqnarray}
q\equiv\textrm{e}^{\alpha/L}\,.
\end{eqnarray}

\medskip 

The remaining task is to check a classical analogue of $q$-Serre relations, 
which are deduced by introducing the classical $q$-Poisson bracket \cite{KMY-QAA},
\begin{eqnarray}
\{J^A,J^B\}_{q\rm P}\equiv \{J^A,J^B\}_{\rm P} + \frac{\alpha}{L} (\beta_A,\beta_B)J^BJ^A\,.
\end{eqnarray}
Here $\beta_A$ are the associated root vectors. 
Now $J^A$ and $J^B$ are $c$-number and commutative, hence the ordering in the second term is irrelevant.  

\medskip 

By a direct computations with the Poisson brackets computed in subsection \ref{expansion:sec}, 
the classical $q$-Serre relations are evaluated as 
\begin{eqnarray}
\{E_i,\{E_i,\{E_i,E_j\}_{q\rm P}\}_{q\rm P}\}_{q\rm P}=\{F_i,\{F_i,\{F_i,F_j\}_{q\rm P}\}_{q\rm P}\}_{q\rm P}=0\quad \textrm{for}\quad|i-j|=1\,.
\end{eqnarray}
The $q$-Serre relations can also be rewritten in terms of 
$Q_{(1)}^{\widehat{\pm}}$ and $\widetilde{Q}_{(1)}^{\widehat{\pm}}$ like 
\begin{eqnarray}
&&\{Q_{(1)}^{\widehat{\pm}},\{Q_{(1)}^{\widehat{\pm}},\{Q_{(1)}^{\widehat{\pm}},\widetilde{Q}_{(1)}^{\widehat{\mp}}\}_{q\rm P}\}_{q\rm P}\}_{q\rm P}=0\,,\\
&&\{\widetilde{Q}_{(1)}^{\widehat{\mp}},\{\widetilde{Q}_{(1)}^{\widehat{\mp}},\{\widetilde{Q}_{(1)}^{\widehat{\mp}},Q_{(1)}^{\widehat{\pm}}\}_{q\rm P}\}_{q\rm P}\}_{q\rm P}=0\,.
\end{eqnarray}
Thus we have shown the classical analogue of quantum affine algebra 
in the sense of Drinfeld's first realization \cite{Drinfeld1}. 
This algebra can be interpreted as the remnant of the relativistic case \cite{KMY-QAA} 
after taking the fast-moving string limit, while the left Yangians are not realized any more.

\subsection{The classical $r$-matrix}

Finally, let us comment on the classical $r$-matrix. 
One can read off the classical $r$-matrix from the following Poisson bracket of $U(t,x;z)$\,, 
\begin{eqnarray}
\left\{U(t,x;z)\stackrel{\otimes}{,}U(t,y;w)\right\}_{\rm P}=\left[r(z-w),U(t,x;z)\otimes 1
+1\otimes U(t,y;w)\right]\delta(x-y)\,. \label{fund-P}
\end{eqnarray}
This bracket has been evaluated by using the Poisson brackets (\ref{Poisson_n})\,.
Note that this bracket does not contain non-ultra local terms in comparison to the relativistic case 
like principal chiral models. 
Hence there is no difficulty to read off the classical $r$-matrix. This is an advantage to consider 
the fast-moving string limit. 

\medskip 

The resulting classical $r$-matrix is of  trigonometric type  \cite{FT},  
\begin{eqnarray}
&&r(z-w)=\dfrac{2}{L}\dfrac{\alpha}{\sinh(z-w)} \Bigl[-\cosh (z-w) T^0\otimes T^0  + T^1\otimes T^1 + T^2\otimes T^2\Bigr]\,,
\end{eqnarray}
and satisfies the classical Yang-Baxter equation, 
\begin{eqnarray}
\left[r_{12}(z-w),r_{13}(z-u)\right]+\left[r_{12}(z-w),r_{23}(w-u)\right] 
+\left[r_{13}(z-u),r_{23}(w-u)\right]=0\,. \label{CYB}
\end{eqnarray}
Thus it ensures that the system is classically integrable.

\section{Integrability of null-like warped $SL(2)$ LLSM}

In this section, we consider a {\it null-like warped $SL(2)$ LLSM}. 
This system is derived as a pp-wave like limit of the time-like warped $SL(2)$ LLSM, 
as shown in detail in Appendix D. The resulting system also coincides with a fast-moving string limit 
of Sch$_3\times$S$^1$ subsector of the Jordanian deformed AdS$_5\times$S$^5$ superstring action 
\cite{KMY-JordanianAdSxS,KMY-JordanianIIB}. 
The associated infinite-dimensional symmetries are also discussed 
by performing non-local gauge transformations which correspond to Jordanian twists.

\subsection{The classical action and Lax pair}

The classical action of the null-like warped $SL(2)$ LLSM is given by \cite{KameYoshi},  
\begin{eqnarray}
S&=&\frac{L}{2}\int^\infty_{-\infty}\!\!\!dtdx\left[
-\frac{C}{2}\left(\cosh\rho+\sin\psi\sinh\rho\right)^2+\left(\cosh\rho-1\right)\partial_t\psi \right. \nonumber \\
&&\left.\hspace{3cm} -\frac{\lambda}{16\pi^2L^2}\left[\left(\partial_x\rho\right)^2+\sinh^2\rho\left(\partial_x\psi\right)^2\right]\right]\,. 
\end{eqnarray}
This system is derived as a pp-wave like limit of the time-like warped $SL(2)$ LLSM, 
as shown in detail in Appendix D. The resulting system also coincides with a fast-moving string limit 
of Sch$_3\times$S$^1$ subsector of the Jordanian deformed AdS$_5\times$S$^5$ superstring action 
\cite{KMY-JordanianAdSxS,KMY-JordanianIIB}. 

\medskip 

It is convenient to introduce a vector notation $n^a$\,, 
\begin{eqnarray}
n^0 = -\cosh\rho\,,\qquad n^1 =\sinh\rho\sin\psi\,, \qquad n^2 =\sinh\rho\cos\psi \,.
\end{eqnarray}
Then the classical equations of motion are written as 
\begin{eqnarray}
&&\partial_t n^0=\frac{\lambda}{8\pi^2L^2}\left(n^1\partial_x^2 n^2-n^2\partial_x^2 n^1\right)-C\left(n^0-n^1\right)n^2\,, \nonumber \\
&&\partial_t n^1=\frac{\lambda}{8\pi^2L^2}\left(n^0\partial_x^2 n^2-n^2\partial_x^2 n^0\right)-C\left(n^0-n^1\right)n^2\,, \nonumber \\
&&\partial_t n^2=\frac{\lambda}{8\pi^2L^2}\left(-n^0\partial_x^2 n^1+n^1\partial_x^2 n^0\right)-C\left(n^0-n^1\right)^2\,. 
\label{eomXXN}
\end{eqnarray}
These are summarized to a simpler form 
\begin{eqnarray}
	\partial_{t} n_a =  \varepsilon_{abc} n^b \Bigl( \dfrac{\lambda}{8\pi^{2}L^{2}}\partial_{x}^{2} n^c + \mathcal{J}^{c}_{~d} n^d\Bigr) 
\qquad (a=0,1,2)\,, 
	\label{eom-XXN}
\end{eqnarray}
with the anisotropic matrix $\mathcal{J}$ 
\begin{eqnarray}
\mathcal{J}^{a}_{~b} =  \begin{pmatrix}
			j + C &-C & 0 \\
			C & j - C & 0 \\
			0 & 0 &  j
			\end{pmatrix}
\qquad (j~: \textrm{an arbitrary const.})\,. 
\end{eqnarray}
The equations (\ref{eom-XXN}) describe null-like deformed Landau-Lifshitz equations. 

\paragraph{Lax pair.}

First of all, it is helpful to introduce the light-cone expressions as 
\begin{eqnarray}
n^{\pm} \equiv \frac{n^0\pm n^1}{\sqrt{2}}\,. 
\end{eqnarray}
Then the Lax pair of the null-like warped $SL(2)$ LLSM is given by 
\begin{eqnarray}
U(t,x;z)&=&\dfrac{\alpha}{z}
\left[-\left(n^+-\frac{Cz^2}{2}n^-\right)T^-
+n^2T^2-n^-T^+\right] \,, \label{Lax-XNN} \\
V(t,x;z)&=&\dfrac{\beta}{z}
\Biggl[
-\left(\left(n^2\partial_x n^+-n^+\partial_x n^2\right)
-\frac{Cz^2}{2}\left(n^-\partial_x n^2-n^2\partial_x n^-\right)\right)T^- \Biggr. \nonumber \\
&&\hspace{1cm}+\Biggl. \left(n^-\partial_xn^+-n^+\partial_xn^-\right)T^2
-\left(n^-\partial_x n^2-n^2\partial_x n^-\right)T^+\Biggr] \nonumber \\
&&+\dfrac{\alpha\beta}{z^2} 
\left[-\left(n^++\frac{Cz^2}{2}n^-\right)T^-+n^2T^2-n^-T^+\right]\,, \nonumber
\end{eqnarray}
where a spectral parameter $z\in\mathbb{C}$ and new parameters have been introduced as 
\begin{eqnarray}
\alpha \equiv \frac{4\pi L}{\sqrt{\lambda}}\,, \qquad
\beta \equiv -\frac{\sqrt{\lambda}}{2\pi L}\,.  
\end{eqnarray}
The Lax pair (\ref{Lax-XNN}) can be derived from the Lax pair of the time-like warped $SL(2)$ LLSM 
by taking a scaling limit, as shown in Appendix D.

\medskip 

The classical equations of motion (\ref{eomXXN}) are reproduced from the commutation relation
\begin{eqnarray}
\Bigl[\partial_t-V(t,x;z),\partial_x-U(t,x;z)\Bigr]=0\,, 
\label{flatness_Lax_XXN}
\end{eqnarray}
and one can check that the Lax pair (\ref{Lax-XNN}) works well.

\subsection{$q$-deformed  Poincar$\acute{\textrm{e}}$ algebras}

Let us consider the symmetry of the null-like warped $SL(2)$ LLSM. 
Due to the deformation, the original $SL(2)$ symmetry is broken to $U(1)$\,. 
As in the previous cases, the broken components are still realized as non-local symmetries. 

\medskip 

The boundary condition is sensitive to the argument on the non-local symmetries. 
In the present case, it is supposed that $n^- $ and $n^2$ vanish at the spatial infinities. 
The detail of the boundary condition is described in Appendix C.

\medskip 
 
The unbroken $U(1)$ charge is constructed as 
\begin{eqnarray}
Q^-=-\frac{L}{2}\int^\infty_{-\infty}\!\!\!dx\,n^-(x)\,.  
\end{eqnarray} 
For the broken components $+$ and $2$\,,  
one can find non-local conserved charges 
\begin{eqnarray}
Q^{+}=-\frac{L}{2} \int^\infty_{-\infty}\!\!\!dx~{\rm e}^{\sqrt{C}\alpha\chi(x)}n^{+}(x)\,,
\qquad Q^{2}=-\frac{L}{2} \int^\infty_{\infty}\!\!\!dx~{\rm e}^{\sqrt{C}\alpha\chi(x)}n^{2}(x)\,.
\end{eqnarray}
Here $\chi(x)$ is a non-local field defined as 
\begin{eqnarray}
\chi(x) \equiv \frac{1}{2}\int^\infty_{-\infty}\!\!\!dy\,\epsilon(x-y)\,n^-(y)\,. 
\end{eqnarray}
The non-local field satisfies the following relations,
\begin{eqnarray}
\partial_x\chi=n^-\,, \qquad 
\partial_t\chi=-\frac{\lambda}{8\pi^2L^2}\left(n^-\partial_x n^2-n^2\partial_x n^-\right)\,.
\end{eqnarray}
These are useful to show the conservation laws of $Q^+$ and $Q^2$\,. 

\medskip 

The next task is to compute the Poisson brackets of $Q^2$ and $Q^{\pm}$\,. 
The Poisson brackets of the dynamical variables $n^a\,(a=2,\pm)$ are 
\begin{eqnarray}
\Big\{n^a(x),n^b(y)\Bigr\}_{\rm P}=-\frac{2}{L}\varepsilon^{ab}_{~~c}n^c(x)\delta(x-y) \,. 
\label{Poisson_n'}
\end{eqnarray} 
With (\ref{Poisson_n'})\,, the Poisson brackets of the charges are evaluated as
\begin{eqnarray}
&&\Bigl\{Q^{+},Q^-\Bigr\}_{\rm P}= - Q^2\,,\nonumber \\
&&\Bigl\{Q^{+},Q^{2}\Bigr\}_{\rm P}=- Q^+ \cosh\biggl(\dfrac{\sqrt{C}\,\alpha}{L}Q^-\biggr)\,, 
\label{Po} \\
&&\Bigl\{Q^{-},Q^{2}\Bigr\}_{\rm P}=\dfrac{L}{\sqrt{C}\,\alpha}
\sinh\biggl(\dfrac{\sqrt{C}\,\alpha}{L}Q^-\biggr)\,. \nonumber
\end{eqnarray}
This is a classical analogue of a non-standard $q$-deformation of $sl(2)$\,, 
where the deformation parameter $q$ is defined as  
\[
q \equiv \textrm{e}^{\frac{\sqrt{C}\,\alpha}{L}}\,.
\] 
The resulting Poisson algebra (\ref{Po}) is isomorphic to a $q$-deformed Poincar$\acute{\textrm{e}}$ algebra 
\cite{q-Poincare,Ohn} with an appropriate rescaling the charges. 

\medskip 

In addition, there exists another set of non-local conserved charges, 
\begin{eqnarray}
\widetilde{Q}^{+}=-\frac{L}{2} \int^\infty_{-\infty}\!\!\!dx~{\rm e}^{-\sqrt{C}\alpha\chi(x)}\,n^{+}(x)\,,\quad 
\widetilde{Q}^{2}=-\frac{L}{2} \int^\infty_{-\infty}\!\!\!dx~{\rm e}^{-\sqrt{C}\alpha\chi(x)}\,n^{2}(x)\,.
\end{eqnarray}
These are obtained by flipping the sign of $\sqrt{C}$ in $Q^+$ and $Q^2$\,.   
By construction, the charges $Q^-\,, \widetilde{Q}^+$ and $\widetilde{Q}^2$ also generate 
another $q$-deformed Poincar$\acute{\textrm{e}}$ algebra.

\medskip 

Note that the mixed Poisson brackets like  $\bigl\{Q^{a},\widetilde{Q}^b\bigr\}_{\rm P}$ 
should be taken into account. Then the algebra is extended to 
an infinite-dimensional symmetry referred to as the {\it exotic} symmetry 
in \cite{exotic}. It is straightforward to reproduce this infinite-dimensional algebra, 
but we will not do that here. 
Instead, we will derive Yangians by undoing Jordanian twists in the next subsection. 

\medskip 

It would be interesting to argue the corresponding classical $r$-matrix. 
From the following Poisson bracket 
\begin{eqnarray}
\left\{U(t,x;z)\stackrel{\otimes}{,}U(t,y;w)\right\}_{\rm P}
=\left[r(z-w),U(t,x;z)\otimes 1+1\otimes U(t,y;w)\right]\delta(x-y)\,.
\end{eqnarray}
the classical $r$-matrix is easily obtained. There is no difficulty of non-ultra local terms again.
The resulting $r$-matrix is deformed like 
\begin{eqnarray}
r(z-w)=\dfrac{2}{L}\Bigl[\dfrac{\alpha}{z-w}\gamma_{ab}T^a\otimes T^b + \frac{C\alpha}{2} (z-w)T^-\otimes T^-\Bigr]\,,   
\end{eqnarray}
but it still satisfies the classical Yang-Baxter equation (\ref{CYB})\,. 
The $r$-matrix of this type is discussed in \cite{Jordanian}.

\subsection{Jordanian twists and $sl(2)$ Yangians}

Let us consider infinite-dimensional symmetries by performing non-local gauge transformations. 
It has been shown in \cite{Jordanian-KMY} that the null-like deformation may be interpreted 
as Jordanian twists in the relativistic case.  
In fact, this interpretation is still applicable in the present non-relativistic case.  
That is, the structure of Jordanian twists remains even after taking the fast-moving string limit. 

\medskip 

First of all, by following \cite{Jordanian-KMY}, let us derive isotropic Lax pairs 
$({\mathcal U}^{(\pm)}(z),{\mathcal V}^{(\pm)}(z))$\,. 
These are obtained from the anisotropic Lax pair (\ref{Lax-XNN})
by performing non-local gauge transformations. 
The derivation of the isotropic Lax pairs are described in Appendix E. 

\medskip 

The resulting isotropic Lax pairs $({\mathcal U}^{(\pm)}(z),{\mathcal V}^{(\pm)}(z))$ are given by 
\begin{eqnarray}
{\mathcal U}^{(+)}(z)&=&\dfrac{\alpha}{z}\left[-{\mathcal N}^-T^++{\mathcal N}^2T^2
-{\mathcal N}^+T^-\right]\,, \\
{\mathcal V}^{(+)}(z)&=&\dfrac{\beta}{z}
\Bigl[-\left({\mathcal N}^-\partial_x {\mathcal N}^2-{\mathcal N}^2\partial_x {\mathcal N}^-\right)T^+
+\left({\mathcal N}^-\partial_x{\mathcal N}^+
-{\mathcal N}^+\partial_x{\mathcal N}^-\right)T^2
 \nonumber \\
&&\hspace{0.6cm}\Bigl.-\left({\mathcal N}^2\partial_x {\mathcal N}^+
-{\mathcal N}^+\partial_x {\mathcal N}^2\right)T^-\Bigr] 
+\dfrac{\alpha\beta}{z^2} 
\Bigl[-{\mathcal N}^-T^++{\mathcal N}^2T^2-{\mathcal N}^+T^-\Bigr]\,, \nonumber \\
{\mathcal U}^{(-)}(z)&=&\dfrac{\alpha}{z}\left[-\widetilde{\mathcal N}^-T^+
+\widetilde{\mathcal N}^2T^2-\widetilde{\mathcal N}^+T^-\right]\,, \nonumber \\
{\mathcal V}^{(-)}(z)&=&\dfrac{\beta}{z}
\Bigl[-\left(\widetilde{\mathcal N}^-\partial_x \widetilde{\mathcal N}^2
-\widetilde{\mathcal N}^2\partial_x \widetilde{\mathcal N}^-\right)T^+
+\left(\widetilde{\mathcal N}^-\partial_x\widetilde{\mathcal N}^+
-\widetilde{\mathcal N}^+\partial_x\widetilde{\mathcal N}^-\right)T^2
 \nonumber \\
&&\hspace{0.6cm}\Bigl.-\left(\widetilde{\mathcal N}^2\partial_x \widetilde{\mathcal N}^+
-\widetilde{\mathcal N}^+\partial_x \widetilde{\mathcal N}^2\right)T^-\Bigr] 
+\dfrac{\alpha\beta}{z^2} 
\Bigl[-\widetilde{\mathcal N}^-T^++\widetilde{\mathcal N}^2T^2
-\widetilde{\mathcal N}^+T^-\Bigr]\,. \nonumber
\end{eqnarray}
Here ${\mathcal N}^a$ and $\widetilde{\mathcal N}^a$ are 
the components of non-local unit-vectors defined in Appendix E. 
Note that there are two kinds of Lax pair according to the choice of the twists 
(non-local gauge transformations). 
The analysis to be performed below is almost irrelevant to the choice. 
Hence we will concentrate on the $(+)$ superscript hereafter. 

\medskip 

Now there is a great advantage because one can follow the standard prescription  
to consider the classical integrable structure. 
With a general prescription, the monodromy matrix is constructed. 
Then, by expanding the monodromy matrix, 
an infinite number of conserved charges are obtained. Here we shall list the first three charges:  
\begin{eqnarray}
&&\mathcal{Q}_{(0)}^a = -\frac{L}{2}\int^\infty_{-\infty}\!\!\!dx~\mathcal{N}^a(x)\,,\notag\\ 
&&\mathcal{Q}_{(1)}^a = \Bigl(-\dfrac{L}{2}\Bigr)^2\!\!\!\int^\infty_{-\infty}\!\!\!dx\int^\infty_{-\infty}\!\!\!dy~\dfrac{1}{4}\epsilon(x-y)\varepsilon_{bc}^{~~a}~\mathcal{N}^b(x)\, \mathcal{N}^c(y)\,,\notag \\
&&\mathcal{Q}_{(2)}^a = \Bigl(-\dfrac{L}{2}\Bigr)^3 \!\!\!\int^\infty_{-\infty}\!\!\!dx\int^\infty_{-\infty}\!\!\!dy\int^\infty_{-\infty}\!\!\!dz~\dfrac{1}{12}\epsilon(x-y) \epsilon(x-z)\gamma_{bc} \nonumber \\ 
&& \hspace*{5cm}  \times \Bigl[\mathcal{N}^a(x)\mathcal{N}^b(y)\mathcal{N}^c(z) 
-\,\mathcal{N}^b(x)\mathcal{N}^a(y)\mathcal{N}^c(z)\Bigr]\,.
\label{yangian}
\end{eqnarray}
Note that these are quite similar to Yangian generators constructed 
from a conserved local current satisfying the flatness condition in the symmetric coset case. 
However, in the present case, the components of $\mathcal{N}^a$ are non-local. 
Hence it is necessary to check the Poisson brackets of them concretely.

\medskip 

The Poisson brackets of ${\mathcal N}^a$ are given by  
\begin{eqnarray}
\Big\{\mathcal{N}^a(x),\mathcal{N}^b(y)\Bigr\}_{\rm P}=-\frac{2}{L}\varepsilon^{ab}_{~~c}~\mathcal{N}^c(x)\delta(x-y)\quad(a=2,\pm)\,. 
\label{Poisson_N}
\end{eqnarray}
With (\ref{Poisson_N}), the Poisson brackets of $\mathcal{Q}_{(0)}^a $ and $\mathcal{Q}_{(1)}^a $  are computed as
\begin{eqnarray}
&&\Bigl\{\mathcal{Q}_{(0)}^{a},\mathcal{Q}_{(0)}^b\Bigr\}_{\rm P}= \varepsilon^{ab}_{~~c} \mathcal{Q}_{(0)}^c\,, \qquad 
\Bigl\{\mathcal{Q}_{(1)}^{a},\mathcal{Q}_{(0)}^b\Bigr\}_{\rm P}= \varepsilon^{ab}_{~~c} \mathcal{Q}_{(1)}^c\,,\notag \\
&&\Bigl\{\mathcal{Q}_{(1)}^{a},\mathcal{Q}_{(1)}^b\Bigr\}_{\rm P}= \varepsilon^{ab}_{~~c}\Bigl[ \mathcal{Q}_{(2)}^c 
- \frac{1}{12}\gamma_{ab}\mathcal{Q}_{(0)}^a \mathcal{Q}_{(0)}^b \mathcal{Q}_{(0)}^c\Bigr]\,.
\label{yangian_alg}
\end{eqnarray}
These are the defining relations of Yangian in the sense of Drinfeld's first realization \cite{Drinfeld1}. 
But the check of the defining relations has not been completed yet. 
The remaining task is to check the Serre relations. Indeed, one can show the following relations,
\begin{eqnarray}
&&\Bigl\{\Bigl\{\mathcal{Q}_{(1)}^{+},\mathcal{Q}_{(1)}^-\Bigr\}_{\rm P}, \mathcal{Q}_{(1)}^{2}\Bigr\}_{\rm P}= \frac{1}{4}\Bigl(\mathcal{Q}_{(1)}^{+} \mathcal{Q}_{(0)}^{-}- \mathcal{Q}_{(0)}^{+}\mathcal{Q}_{(1)}^{-}\Bigr)\mathcal{Q}_{(0)}^{2}\,,\notag \\  
&&\Bigl\{\Bigl\{\mathcal{Q}_{(1)}^{2},\mathcal{Q}_{(1)}^{\pm}\Bigr\}_{\rm P}, \mathcal{Q}_{(1)}^{\pm}\Bigr\}_{\rm P}= \frac{1}{4}\Bigl(\mathcal{Q}_{(1)}^{2} \mathcal{Q}_{(0)}^{\pm}- \mathcal{Q}_{(0)}^{2}\mathcal{Q}_{(1)}^{\pm}\Bigr)\mathcal{Q}_{(0)}^{\pm}\,,\notag \\
&&\Bigl\{\Bigl\{\mathcal{Q}_{(1)}^{+},\mathcal{Q}_{(1)}^-\Bigr\}_{\rm P}, \mathcal{Q}_{(1)}^{\pm}\Bigr\}_{\rm P}\mp\Bigl\{\Bigl\{\mathcal{Q}_{(1)}^{2},\mathcal{Q}_{(1)}^{\pm}\Bigr\}_{\rm P}, \mathcal{Q}_{(1)}^{2}\Bigr\}_{\rm P}\notag \\= &&\frac{1}{4}\Bigl(\mathcal{Q}_{(1)}^{+} \mathcal{Q}_{(0)}^{-}- \mathcal{Q}_{(0)}^{+}\mathcal{Q}_{(1)}^{-}\Bigr)\mathcal{Q}_{(0)}^{\pm} \mp\frac{1}{4}\Bigl(\mathcal{Q}_{(1)}^{2} \mathcal{Q}_{(0)}^{\pm}- \mathcal{Q}_{(0)}^{2}\mathcal{Q}_{(1)}^{\pm}\Bigr)\mathcal{Q}_{(0)}^{2}\,, 
\end{eqnarray}
and the Serre relations are also satisfied. Note again that there is no non-ultra local term. 
Thus the Yangian algebra ${\mathcal Y}(sl(2))$ is generated in a well-defined manner.

\medskip 

Starting from the $(-)$ superscript starting from ${\mathcal U}^{(-)}(z)$ and ${\mathcal V}^{(-)}(z)$\,, 
one can derive the same result. 
That is, the same Poisson brackets are derived, up to the replacement 
of $\mathcal{Q}_{(n)}^{a}$ by $\widetilde{\mathcal{Q}}_{(n)}^{a}$\,, 
while $\widetilde{\mathcal{Q}}_{(n)}^{a}$ are not identical with  $\mathcal{Q}_{(n)}^{a}$\,. 

\medskip 

Finally let us comment on the classical $r$-matrix. 
From the Poisson brackets
\begin{eqnarray}
&& \left\{{\mathcal U}^{(\pm)}(t,x;z)\stackrel{\otimes}{,}{\mathcal U}^{(\pm)}(t,y;w)\right\}_{\rm P} \nonumber \\ 
&=& \left[r^{(\pm)}(z-w),~ {\mathcal U}^{(\pm)}((t,x;z)\otimes 1+1\otimes {\mathcal U}^{(\pm)}(t,y;w)\right]\delta(x-y)\,, \notag
\end{eqnarray}
one can read off the classical $r$-matrices. 
These are Yang's $r$-matrix of  rational type,  
\begin{eqnarray}
&&r^{(\pm)}(z-w)=\dfrac{2}{L}\dfrac{\alpha}{z-w}\gamma_{ab}T^a\otimes T^b \,,
\end{eqnarray}
and satisfy the classical Yang-Baxter equation (\ref{CYB})\,. 
Note that it is independent of $C$\,. 

\subsection{The relation between $q$-Poincar$\acute{\textrm{e}}$ algebras and Yangians}

It is worth listing the relation between the exotic symmetry and the Yangians. 

\medskip 

For simplicity, let us concentrate on one of the Yangians generated by 
$\mathcal{Q}^{\pm}$ and $\mathcal{Q}^2$\,. The similar argument holds also for 
the other Yangian.  
The Yangian charges can be represented in terms of the $q$-deformed Poincar$\acute{\textrm{e}}$ 
charges $Q^a$ and $\widetilde{Q}^{a}$\,. 

\medskip 

The level-zero Yangian charges are expressed as   
\begin{eqnarray}
&&\mathcal{Q}_{(0)}^{+}= \textrm{e}^{\frac{\sqrt{C}\alpha}{L}Q^-}Q^+ 
+\frac{\sqrt{C}\alpha}{L}\Bigl(\textrm{e}^{\frac{\sqrt{C}\alpha}{L}Q^-}Q^2 \Bigr)^2\,,\notag \\
&&\mathcal{Q}_{(0)}^{2}=\textrm{e}^{\frac{\sqrt{C}\alpha}{L}Q^-}Q^2\,,\notag \\
&&\mathcal{Q}_{(0)}^{-}=\frac{1}{2\frac{\sqrt{C}\alpha}{L}}
\Bigl(1-\textrm{e}^{-2\frac{\sqrt{C}\alpha}{L}Q^-}\Bigr)\,.
\end{eqnarray}
Then the level-one charges are 
\begin{eqnarray}
&&\mathcal{Q}_{(1)}^{+}=\frac{1}{4}\biggl[\frac{1}{\frac{\sqrt{C}\alpha}{L}} 
\Bigl\{\Bigl\{\widetilde{Q}^{2},Q^+\Bigr\}_{\rm P}, Q^{+}\Bigr\}_{\rm P} 
+ \Bigl\{\widetilde{Q}^{2},Q^+\Bigr\}_{\rm P}Q^2 \notag\\
&&\quad\quad\quad\quad\quad- \Bigl(\sinh\Bigl(\frac{\sqrt{C}\alpha}{L}Q^-\Bigr)\widetilde{Q}^{2} 
+\cosh\Bigl(\frac{\sqrt{C}\alpha}{L}Q^-\Bigr)Q^2\Bigr)Q^+\biggr]\,,\notag \\
&&\mathcal{Q}_{(1)}^{2}=\frac{1}{4\frac{\sqrt{C}\alpha}{L}}\Bigl(\Bigl\{\widetilde{Q}^{2},Q^+\Bigr\}_{\rm P}
-\cosh\Bigl(\frac{\sqrt{C}\alpha}{L}Q^-\Bigr)Q^+\Bigr) +\frac{1}{4}Q^2\Bigl(\widetilde{Q}^2 
- Q^2\Bigr)\,,\notag \\
&&\mathcal{Q}_{(1)}^{-}=\frac{1}{4\frac{\sqrt{C}\alpha}{L}}\Bigl(\widetilde{Q}^2 - Q^2\Bigr)\,.
\end{eqnarray}
These relations are the same as the ones in the relativistic case \cite{Jordanian-KMY}. 
That is, this structure survives the fast-moving string limit. 

\subsection{A possible relation between gravitational solutions}

Finally, we should comment on a possible relation between the $q$-deformed AdS$_5\times$S$^5$ 
and a Jordanian deformation of AdS$_5\times$S$^5$ argued in \cite{KMY-JordanianIIB}. 

\begin{figure}[t]
\begin{center}
\includegraphics[scale=.45]{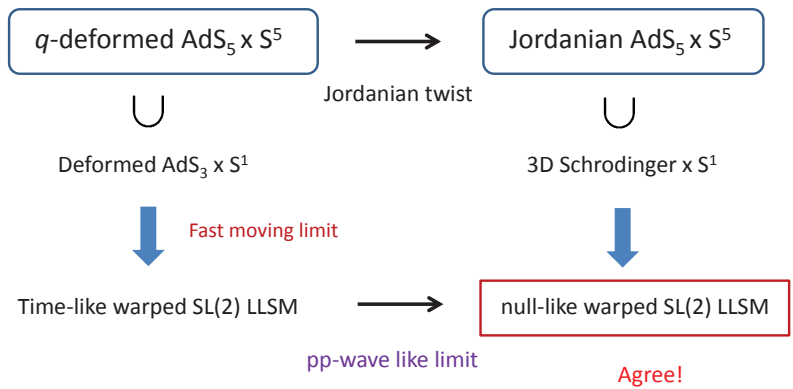}
\end{center}
\vspace*{-0.7cm}
\caption{A possible relation between a Jordanian twist and a pp-wave like limit.}
\label{1:fig}
\end{figure}

The null-like warped $SL(2)$ LLSM is obtained 
as a pp-wave like limit of the time-like warped one, 
as explained in Appendix D. 
The identical null-like 
warped $SL(2)$ LLSM is also derived from a string sigma model on Sch$_3\times$S$^1$\,.  
Then the geometry of Sch$_3\times$S$^1$ is contained as a subspace of 
a Jordanian deformed AdS$_5\times$S$^5$ \cite{KMY-JordanianIIB}. 

\medskip 

Thus one may expect a relation between the $q$-deformed AdS$_5\times$S$^5$ 
and the Jordanian deformed AdS$_5\times$S$^5$\,. Indeed, this is the case. 
As explained in detail in \cite{KMY-JordanianAdSxS}\,, 
the $r$-matrix that leads to the Jordanian deformed solution \cite{KMY-JordanianIIB} 
is constructed by performing a Jordanian twist of the $r$-matrix of Drinfeld-Jimbo type 
\cite{DMV-string,ABF}. Hence this observation suggests that 
the Jordanian twist at the $r$-matrix level corresponds to the pp-wave like limit 
at the geometry level, as depicted in Figure~\ref{1:fig}. 
It would be an interesting issue to make this correspondence more precise.

\section{Conclusion and Discussion}

In this paper, we have derived anisotropic LLSMs from bosonic subsectors 
of the $q$-deformed AdS$_5\times$S$^5$ superstring action 
by taking fast-moving string limits. Then we have investigated the classical integrability 
of the LLSMs from the viewpoint of infinite-dimensional symmetries. 

\medskip 

Concretely speaking, we have considered the subsectors, 1) deformed AdS$_3\,\times\,$S$^1$ and 
2) R\,$\times$\,deformed S$^3$\,. 
By taking fast-moving string limits, a time-like warped $SL(2)$ LLSM and a squashed S$^3$ LLSM 
have been derived for the case 1) and the case 2), respectively. It is remarkable that the resulting LLSMs 
coincide precisely with the ones obtained from a time-like warped AdS space and a squashed S$^3$ 
\cite{KameYoshi}.  
 
\medskip 
 
Then infinite-dimensional symmetries have been revealed under an appropriate boundary condition.  
In the case 1), a quantum affine algebra $U_q(\widehat{sl(2)})$ has been shown explicitly 
by computing the Poisson brackets of conserved non-local charges. 
In the case 2), a quantum affine algebra $U_q(\widehat{su(2)})$ is realized. 
It should be noted that non-ultra local terms do not appear 
in computing the Poisson algebra and hence there is no ambiguity 
in studying the classical integrable structure, in comparison to principal chiral models. 

\medskip 

For the case 1)\,,  a pp-wave like limit has been applied. 
The resulting system coincides with a null-like warped $SL(2)$ LLSM obtained as 
a fast-moving string limit of a string sigma model on Sch$_3\times$S$^1$ \cite{KameYoshi}.  
As a result, a couple of Yangians ${\mathcal Y}(sl(2))$ have been revealed 
by performing non-local gauge transformations which correspond to undoing Jordanian twists. 
In addition, we have argued a possible relation between the $q$-deformed AdS$_5\times$S$^5$ 
and Jordanian deformed AdS$_5\times$S$^5$\,.  

\medskip

It is interesting to consider some generalizations of our result. There would be various directions. 
The first is to consider a long-range generalization . By following the works \cite{KRT,KT,MTT}, 
it would be possible to argue a long-range generalization of our result 
(For example, for a long-range generalization of the XXZ model see \cite{BFdLL}). 
The next is to consider larger subsectors, 
or directly the full sector (For a supersymmetric generalization of the undeformed case, see 
\cite{Rafael2,Stefanski}). It would also be nice to consider the LLSMs obtained here at the quantum level, 
for example, by following \cite{RTT,Tirziu,quantumLL}.  

\medskip 

There are many open problems concerning anisotropic LLSMs. 
We hope that our result would open up a new arena to study $q$-deformations 
of the AdS$_5\times$S$^5$ superstring. 

\subsection*{Acknowledgments}

We are grateful to Io Kawaguchi for helpful discussions and collaborations at an earlier
stage. We  also appreciate  Niklas Beisert, Marius de Leeuw, Sergey Frolov, Takuya Matsumoto and Matthias Staudacher for useful discussions. 
The work of TK was supported by the Japan Society for the Promotion of Science (JSPS). 

\appendix 

\section{Our convention of  generators}

We shall summarize here our conventions of the $sl(2)$ and $su(2)$ generators.

\subsection{The $sl(2)$ generators}
The $sl(2)$ generators $T^a$'s are defined as 
\begin{eqnarray}
T^0 \equiv \frac{i}{2}\sigma_2\,, \quad
T^1 \equiv \frac{1}{2}\sigma_1\,, \quad
T^2 \equiv \frac{1}{2}\sigma_3\,, 
\end{eqnarray}
where $\sigma_i~~(i=1,2,3)$ are the standard Pauli matrices. 
Then the commutation relation and the normalization are  
\begin{eqnarray}
\left[T^a,T^b\right]=\varepsilon^{ab}_{~~c}T^c\,, \qquad 
{\rm Tr}\left(T^aT^b\right)=\frac{1}{2}\gamma^{ab}\,,
\end{eqnarray}
with a totally anti-symmetric tensor $\varepsilon^{ab}_{~~c}$ normalized as 
$\varepsilon^{01}_{~~2}=+1$\,. The components of $\gamma^{ab}$ are given by  
\begin{eqnarray}
-\gamma^{00}=\gamma^{11}=\gamma^{22}=+1\,, \qquad
\gamma^{ab}=0~~(a\neq b)\,. 
\end{eqnarray}
The $sl(2)$ indices are raised and lowered by $\gamma^{ab}$ and its inverse, respectively.  

\medskip 

It is convenient to introduce $T^\pm$ defined as 
\begin{eqnarray}
T^\pm \equiv \frac{1}{\sqrt{2}}\left(T^0\pm T^1\right)\,, 
\end{eqnarray}
where the components of $\varepsilon^{abc}$ and $\gamma^{ab}$ are 
\begin{eqnarray}
&&\varepsilon^{-+2}=+1\,, \qquad 
-\gamma^{-+}=-\gamma^{+-}=\gamma^{22}=+1\,. \nonumber
\end{eqnarray}
It is also helpful to introduce other linear combinations $T^{\widehat{\pm}}$ defined as 
\begin{eqnarray}
T^{\widehat{\pm}}\equiv \frac{1}{\sqrt{2}}\left(T^1\pm iT^2\right)\,,
\end{eqnarray}
The commutation relation and the normalization are  
\begin{eqnarray}
\left[T^a,T^b\right]=i\,\varepsilon^{ab}_{~~c}T^c\,, \qquad 
{\rm Tr}\left(T^aT^b\right)=\frac{1}{2}\gamma^{ab}\,,
\end{eqnarray}
where the components of $\varepsilon^{abc}$ and $\gamma^{ab}$ are expressed as 
\begin{eqnarray}
&&\varepsilon^{0{\widehat{-}} {\widehat{+}}}=+1\,, \qquad 
-\gamma^{00}=\gamma^{\widehat{-}\widehat{+}}=\gamma^{\widehat{+}\widehat{-}}=+1\,. \nonumber
\end{eqnarray}

\subsection{The $su(2)$ generators}
The $su(2)$ generators $T^a$'s are defined as 
\begin{eqnarray}
T^1 \equiv \frac{1}{2i}\sigma_1\,, \quad
T^2 \equiv \frac{1}{2i}\sigma_2\,, \quad
T^3 \equiv \frac{1}{2i}\sigma_3\,, 
\end{eqnarray}
where $\sigma_i~~(i=1,2,3)$ are the standard Pauli matrices. 
Then the commutation relation and the normalization are given by 
\begin{eqnarray}
\left[T^a,T^b\right]=\varepsilon^{ab}_{~~c}T^c\,, \qquad 
{\rm Tr}\left(T^aT^b\right)=\frac{1}{2}\delta^{ab}\,,
\end{eqnarray}
with a totally anti-symmetric tensor $\varepsilon^{ab}_{~~c}$ normalized as 
$\varepsilon^{12}_{~~3}=+1$\,. 
The $su(2)$ indices are raised and lowered by $\delta^{ab}$ and its inverse, respectively.  

\medskip 

It is convenient to introduce $T^\pm$ defined as 
\begin{eqnarray}
T^\pm \equiv \frac{1}{\sqrt{2}}\left(T^1\pm iT^2\right)\,,
\end{eqnarray}
where the components of $\varepsilon^{abc}$ and $\delta^{ab}$ are given by 
\begin{eqnarray}
&&\varepsilon^{-+3}=+1\,, \qquad 
\delta^{33}=\delta^{{+}{-}}=\delta^{{-}{+}}=+1\,. \nonumber
\end{eqnarray}

\section{Integrability of squashed S$^3$ LLSM}

Here we consider the classical integrability of the squashed S$^3$ LLSM and 
the associated infinite-dimensional symmetry. 

\subsection{The classical action and Lax pair}

The classical action of the squashed S$^3$ LLSM is given by 
\begin{eqnarray}
S&=&\frac{L}{2}\int^\infty_{-\infty}\!\!\!dtdx\,\biggl[C\cos^2\theta-\cosh\theta\partial_t\phi  \nonumber \\ 
&& \hspace*{3.0cm}
-\frac{\lambda}{16\pi^2L^2}\left[\left(\partial_x\theta\right)^2+\sin^2\theta\left(\partial_x\phi\right)^2\right]\biggr]\,. 
\label{action_s3}
\end{eqnarray}
When $C=0$\,, the isotropic $S^3$ LLSM is reproduced. 

\medskip 

It is convenient to introduce a vector representation with $n^a$\,.  
The components are 
\begin{eqnarray}
n^1=\sin\theta\sin\phi\,, \qquad
n^2=\sin\theta\cos\phi\,,\qquad
n^3=\cos\theta\,,
\end{eqnarray}
and satisfy a constraint condition,
\begin{eqnarray}
\left(n^1\right)^2+\left(n^2\right)^2+\left(n^3\right)^2=1\,.
\end{eqnarray}
Then the classical equations of motion are rewritten as 
\begin{eqnarray}
&&\partial_t n^1=\frac{\lambda}{8\pi^2L^2}\left(n^2\partial_x^2 n^3-n^3\partial_x^2 n^2\right)+2Cn^2n^3\,, 
\nonumber \\
&&\partial_t n^2=\frac{\lambda}{8\pi^2L^2}\left(n^3\partial_x^2 n^1-n^1\partial_x^2 n^3\right)-2Cn^1n^3\,, 
\nonumber \\
&&\partial_t n^3=\frac{\lambda}{8\pi^2L^2}\left(n^1\partial_x^2 n^2-n^2\partial_x^2 n^1\right)\,.
\label{sqs3eom}
\end{eqnarray}
These are equivalent to the Landau-Lifshitz equations\footnote{Here we have introduced 
the totally anti-symmetric tensor $\varepsilon_{abc}$ with $\varepsilon_{123} = + 1$\,. 
}
\begin{eqnarray}
	\partial_{t} n_a =  \varepsilon_{abc} n^b \left( 
\dfrac{\lambda}{8\pi^{2}L^{2}}\partial_{x}^{2} n^c + \mathcal{J}^{c}_{~d} n^d\right)\qquad (a=1,2,3)\,,
\end{eqnarray}
with an anisotropic matrix $\mathcal{J}$ 
\begin{eqnarray}
\mathcal{J}^{a}_{~b} =  \textrm{diag}( j , j, j+2C) \qquad (j :~\textrm{an~arbitrary~const.})\,.
\end{eqnarray}

\subsubsection*{Lax pair and monodromy matrix} 
Let us introduce the Lax pair of the squashed S$^3$ LLSM \cite{FT}, 
\begin{eqnarray}
U(t,x;z)&=&\dfrac{\alpha}{\sin z} \left[n^1T^1+ n^2T^2 +\cos z\, n^3T^3 \right] \,, \\
V(t,x;z)&=&\dfrac{\beta}{\sin z} \left[\left(n^2\partial_x n^3-n^3\partial_x n^2\right) T^1+\left(n^3\partial_x n^1-n^1\partial_x n^3\right)T^2 \right. \nonumber \\
&&\hspace{1.5cm}\left.+\cos z\left(n^1\partial_x n^2-n^2\partial_xn^1\right)T^3 \right] \nonumber \\
&&-\dfrac{\alpha\beta}{\sin^2 z} \left[\cos z\,n^1T^1+\cos z\,n^2T^2 + n^3T^3\right]\,, \notag 
\end{eqnarray}
where we have introduced a spectral parameter $z\in \mathbb{C}$ and new parameters,
\begin{eqnarray}
\alpha \equiv \dfrac{4\pi L}{\sqrt{\lambda}}\sqrt{C}\,, \qquad 
\beta \equiv \dfrac{\sqrt{\lambda}}{2\pi L} \sqrt{C}\,.
\end{eqnarray}
It is an easy task to check that 
the equations of motion (\ref{sqs3eom}) are reproduced from the commutation relation 
\begin{eqnarray}
\Bigl[\partial_t-V(t,x;z),\partial_x-U(t,x;z)\Bigr]=0\,. 
\label{flatness_Lax_su2}
\end{eqnarray}

The monodromy matrix can be introduced as 
\begin{eqnarray}
M(z) \equiv {\rm P}\exp\left[\int^\infty_{-\infty}\!\!\!dx~U(t,x;z)\right]\,, 
\end{eqnarray}
where $P$ denotes the parth-ordering. This is a conserved quantity again.
As we will see later, the expansions around $z=\pm\infty$ lead to 
a quantum affine algebra $U_q(\widehat{su(2)})$\,.

\subsection{The standard $q$-deformation of $su(2)$}

Let us consider a $q$-deformation of $su(2)$ realized in the squashed S$^3$ LLSM. 

\medskip 

The $SU(2)$ symmetry of the isotropic S$^3$ LLSM is broken to $U(1)$ 
due to the deformation. The remaining $U(1)$ generator is given by  
\begin{eqnarray}
Q^3=-\dfrac{L}{2}\int^\infty_{-\infty}\!\!\!dx~n^3(x)\,.  
\end{eqnarray}
Note here that the broken components of $SU(2)$ are still realized in a non-local way 
even when $C\neq 0$\,, as in the time-like warped $SL(2)$ LLSM. 

\medskip 

To find out the corresponding non-local charges, it is helpful to introduce $n^{{\pm}}$ defined as 
\begin{eqnarray}
n^{{\pm}} \equiv \frac{n^1 \pm i n^2}{\sqrt{2}}\,. 
\end{eqnarray}
We take the rapidly damping condition so that $n^{{\pm}} $ vanish 
at the spatial infinities. The non-local charges are conserved under this boundary 
condition (See Appendix C).

\medskip

The non-local conserved charges are given by 
\begin{eqnarray}
Q^{{\pm}}=-\dfrac{L}{2}\int^\infty_{-\infty}\!\!\!dx~{\rm e}^{\alpha\chi(x)}n^{{\pm}}(x)\,, 
\end{eqnarray}
where $\chi(x)$ is a non-local field defined as 
\begin{eqnarray}
\chi(x)\equiv \frac{1}{2}\int^\infty_{-\infty}\!\!\!dy\,\epsilon(x-y)\,n^3(y)\,.
\end{eqnarray}
Here the following relations are useful to show the conservation laws of $Q^{{\pm}}$\,, 
\begin{eqnarray}
\partial_x\chi=n^3\,, \qquad
\partial_t\chi=i\frac{\lambda}{8\pi^2L^2}\left(n^{{+}}\partial_x n^{{-}}-n^{{-}}\partial_x n^{{+}}\right)\,. 
\end{eqnarray}

\medskip

The next is to compute the Poisson brackets of $Q^3$ and $Q^{{\pm}}$\,. 
The brackets for $n^a\, (a=3,{\pm})$ are 
\begin{eqnarray}
\Big\{n^a(x),n^b(y)\Bigr\}_{\rm P}=-i\frac{2}{L}\,\varepsilon^{ab}_{~~c}n^c(x)\delta(x-y)\,. 
\label{Poisson_su2}
\end{eqnarray}
Non-ultra local terms are not contained again.

\medskip 

With the brackets (\ref{Poisson_su2})\,, 
the brackets of $Q^0$ and $Q^{{\pm}}$ can be evaluated as
\begin{eqnarray}
&\Bigl\{Q^{{\pm}},Q^{3}\Bigr\}_{\rm P}&=\pm iQ^{{\pm}}\,,\\
&\Bigl\{Q^{{+}},Q^{{-}}\Bigr\}_{\rm P}&=- i\frac{L}{2\alpha}\sinh\Bigl(\frac{2\alpha}{L} Q^3\Bigr)\,.\nonumber
\end{eqnarray}
This is a classical analogue of the standard $q$-deformation of $su(2)$ \cite{Drinfeld2,Jimbo}. 

\medskip 

In addition, there exists another set of non-local conserved charges, 
\begin{eqnarray}
\widetilde{Q}^{{\pm}}=-\dfrac{L}{2}\int^\infty_{-\infty}\!\!\!dx~{\rm e}^{-\alpha\chi(x)}n^{{\pm}}(x)\,. 
\end{eqnarray}
These are obtained by changing the sign of $\alpha$ in $Q^{{\pm}}$\,. 
Hence, by construction, $Q^3$ and $\widetilde{Q}^{{\pm}}$ also generate another $q$-deformed $su(2)$ algebra,  
\begin{eqnarray}
&&\Bigl\{\widetilde{Q}^{{\pm}},{Q}^3\Bigr\}_{\rm P}=\pm i\widetilde{Q}^{{\pm}}\,,\\
&&\Bigl\{\widetilde{Q}^{{+}},\widetilde{Q}^{{-}}\Bigr\}_{\rm P}
=-i \dfrac{L}{2\alpha} \sinh\Bigl(\dfrac{2\alpha}{L}Q^3\Bigr)\,.\nonumber
\end{eqnarray}
It is possible to show that a quantum affine algebra 
is generated by $Q^3$\,, $Q^{{\pm}}$ and $\widetilde{Q}^{{\pm}}$ 
in the sense of Drinfeld's first realization \cite{Drinfeld1}. 
The derivation is essentially the same as in the time-like warped  LLSM, 
hence we will not give the derivation.

\subsection{The classical $r$-matrix}

Finally, let us discuss the classical $r$-matrix from (\ref{fund-P}). 
The resulting classical $r$-matrix is of  trigonometric type  \cite{FT},  
\begin{eqnarray}
&&r(z-w)=\dfrac{2}{L}\dfrac{\alpha}{\sin(z-w)} \Bigl[T^1\otimes T^1 + T^2\otimes T^2 + \cos (z-w) T^3\otimes T^3 \Bigr]\,,
\end{eqnarray}
and satisfies the classical Yang-Baxter equation (\ref{CYB})\,.  
This ensures the classical integrability of the squashed S$^3$ LLSM.

\section{Boundary conditions and conservation laws}

We argue here the relation between boundary conditions and the conservation laws 
of non-local charges for the time-like warped $SL(2)$ LLSM, the squashed S$^3$ LLSM 
and the null-like warped $SL(2)$ LLSM. 

\subsection{Time-like warped $SL(2)$ LLSM}

Let us first consider the time-like warped $SL(2)$ LLSM. 

\medskip 

We impose the following boundary condition:
\begin{eqnarray}
n^{\widehat{\pm}}(x)\rightarrow 0 \qquad
\textrm{as} \qquad x\rightarrow \pm \infty\,.
\label{condtime}
\end{eqnarray}
In terms of the coordinates, the boundary condition (\ref{condtime}) is expressed as
\begin{eqnarray}
\rho(x)\rightarrow
0\,,\quad\psi(x)\rightarrow\textrm{const.}\quad
\textrm{as} \quad x\rightarrow \pm \infty\,.
\end{eqnarray}
By using the relation
\begin{eqnarray}
\left(n^0\right)^2-2n^{\widehat{+}}n^{\widehat{-}}=1\,,
\end{eqnarray}
the condition (\ref{condtime}) implies that $(n^0)^2 \to 1$ as $x  \to \pm\infty$\,. 
Because we are working with the parametrization $n^0 = -\cosh\rho$\,, 
the boundary condition for $n^0$ is fixed as follows: 
\begin{eqnarray}
n^0(x)\rightarrow -1 \qquad
\textrm{as} \qquad x\rightarrow \pm \infty\,. \label{bc1}
\end{eqnarray}
It should be noted here that the local charge $Q^0$ is not finite and diverges 
under the condition (\ref{bc1}). 
However, this divergence may have a physical interpretation. 
The value of $Q^0$ is closely related to the length of the spatial direction $x$ of 
the string world-sheet due to the spin chain interpretation. 
We are now considering the infinite spatial direction of the string world-sheet, 
hence the divergence of $Q^0$ may be interpreted as a rather natural result. 

\medskip

Let us next check the conservation laws of $Q^{\widehat{\pm}}$\,.
One can show them as follows:
\begin{eqnarray}
\partial_tQ^{\widehat{+}}&=&-\frac{L}{2} \int^\infty_{-\infty}\!\!\!dx~
\partial_t[{\rm e}^{\alpha\chi(x)}n^{\widehat{+}}(x)] \nonumber\\
&=&-i\frac{\lambda}{16\pi^2L}\int^\infty_{-\infty}\!\!\!dx~
\partial_x[{\rm e}^{\alpha\chi}(n^{\widehat{+}}\partial_xn^0
-n^0\partial_xn^{\widehat{+}})]-i\frac{\sqrt{\lambda}}{4\pi}\sqrt{C}
\int^\infty_{-\infty}\!\!\!dx~\partial_x[{\rm e}^{\alpha\chi}n^{\widehat{+}}]
\notag\\&=&0\,,\notag\\
\label{partial_tQ+}
\partial_tQ^{\widehat{-}}&=&-\frac{L}{2} \int^\infty_{-\infty}\!\!\!dx~
\partial_t[{\rm e}^{\alpha\chi(x)}n^{\widehat{-}}(x)] \nonumber \\
&=&-i\frac{\lambda}{16\pi^2L}\int^\infty_{-\infty}\!\!\!dx~
\partial_x[{\rm e}^{\alpha\chi}(n^0\partial_xn^{\widehat{-}}
-n^{\widehat{-}}\partial_xn^0)]+i\frac{\sqrt{\lambda}}{4\pi}\sqrt{C}
\int^\infty_{-\infty}\!\!\!dx~\partial_x[{\rm e}^{\alpha\chi}n^{\widehat{-}}]\notag\\
&=&0\,.\notag
\label{partial_tQ-}
\end{eqnarray}
Under the condition (\ref{condtime}),  all of the surface terms 
have vanished. 

\medskip 

Similarly, for another set of non-local charges $\widetilde{Q}^{\widehat{\pm}}$\,, 
the conservation laws are shown as 
\begin{eqnarray}
\partial_t\widetilde{Q}^{\widehat{+}}&=&-\frac{L}{2} \int^\infty_{-\infty}\!\!\!dx~
\partial_t[{\rm e}^{-\alpha\chi(x)}n^{\widehat{+}}(x)] \nonumber \\
&=&-i\frac{\lambda}{16\pi^2L}\int^\infty_{-\infty}\!\!\!dx~\partial_x[{\rm e}^{-\alpha\chi}
(n^{\widehat{+}}\partial_xn^0-n^0\partial_xn^{\widehat{+}})]+i\frac{\sqrt{\lambda}}{4\pi}
\sqrt{C}\int^\infty_{-\infty}\!\!\!dx~\partial_x[{\rm e}^{-\alpha\chi}n^{\widehat{+}}]
\notag\\&=&0\,,\notag\\
\label{partial_ttQ+}
\partial_t\widetilde{Q}^{\widehat{-}}&=&-\frac{L}{2} \int^\infty_{-\infty}\!\!\!dx~
\partial_t[{\rm e}^{-\alpha\chi(x)}n^{\widehat{-}}(x)] \nonumber \\
&=&-i\frac{\lambda}{16\pi^2L}\int^\infty_{-\infty}\!\!\!dx~\partial_x[{\rm e}^{-\alpha\chi}
(n^0\partial_xn^{\widehat{-}}-n^{\widehat{-}}\partial_xn^0)]-i\frac{\sqrt{\lambda}}{4\pi}
\sqrt{C}\int^\infty_{-\infty}\!\!\!dx~\partial_x[{\rm e}^{-\alpha\chi}n^{\widehat{-}}]
\notag\\&=&0\,.\notag \label{partial_ttQ-}
\end{eqnarray}
Thus $\widetilde{Q}^{\widehat{\pm}}$ are also conserved under the condition (\ref{condtime}). 

\medskip 

Note that all of the non-local charges are finite under the condition (\ref{condtime}), 
while the local charge $Q^0$ diverges.

\subsection{Squashed S$^3$ LLSM}

Next, we consider the squashed S$^3$ LLSM. 
The argument here is essentially the same as in the case 
of the time-like warped $SL(2)$ LLSM. 

\medskip 

First of all, let us impose the following boundary condition:
\begin{eqnarray}
n^{{\pm}}(x)\rightarrow 0 \qquad
\textrm{as} \qquad x\rightarrow \pm \infty\,.
\label{condsu2}
\end{eqnarray}
In terms of the coordinates, the condition (\ref{condsu2}) is expressed as
\begin{eqnarray}
\theta(x)\rightarrow
0\,,\quad\phi(x)\rightarrow\textrm{const.}\quad
\textrm{as} \quad x\rightarrow \pm \infty\,.
\end{eqnarray}
By using the relation
\begin{eqnarray}
\left(n^3\right)^2+2n^{{+}}n^{{-}}=1\,,
\end{eqnarray}
the condition (\ref{condsu2}) indicates that $(n^3)^2\to 1$ as $x\to\pm\infty$\,. 
Due to the parametrization $n^3 = \cos\theta$\,, the boundary condition 
for $n^3$ is fixed as follows: 
\begin{eqnarray}
n^3(x)\rightarrow 1 \qquad
\textrm{as} \qquad x\rightarrow \pm \infty\,.
\end{eqnarray}
As a result, the local charge $Q^3$ is not finite but diverges again. 
The divergence may also be interpreted physically, as mentioned before. 

\medskip

The conservation laws of $Q^{{\pm}}$ are shown as  
\begin{eqnarray}
\partial_tQ^{{+}}&=&-\frac{L}{2} \int^\infty_{-\infty}\!\!\!dx~
\partial_t[{\rm e}^{\alpha\chi(x)}n^{{+}}(x)] \nonumber \\
&=&-i\frac{\lambda}{16\pi^2L}\int^\infty_{-\infty}\!\!\!dx~\partial_x[{\rm e}^{\alpha\chi}
(n^3\partial_xn^{{+}}-n^{{+}}\partial_xn^3)]+i\frac{\sqrt{\lambda}}{4\pi}
\sqrt{C}\int^\infty_{-\infty}\!\!\!dx~\partial_x[{\rm e}^{\alpha\chi}n^{{+}}]\notag\\&=&0\,,\notag\\
\label{partial_tQ+_su2}
\partial_tQ^{{-}}&=&-\frac{L}{2} \int^\infty_{-\infty}\!\!\!dx~\partial_t[{\rm e}^{\alpha\chi(x)}n^{{-}}(x)] 
\nonumber \\
&=&-i\frac{\lambda}{16\pi^2L}\int^\infty_{-\infty}\!\!\!dx~\partial_x[{\rm e}^{\alpha\chi}
(n^{{-}}\partial_xn^3-n^3\partial_xn^{{-}})]-i\frac{\sqrt{\lambda}}{4\pi}\sqrt{C}\int^\infty_{-\infty}
\!\!\!dx~\partial_x[{\rm e}^{\alpha\chi}n^{-}]\notag\\&=&0\,.\notag
\label{partial_tQ-_su2}
\end{eqnarray}
All of the surface terms 
have vanished under the condition (\ref{condsu2}). 

\medskip 

Also for $\widetilde{Q}^{\pm}$\,, the conservation laws are shown as follows:  
\begin{eqnarray}
\partial_t\widetilde{Q}^{{+}}&=&-\frac{L}{2} \int^\infty_{-\infty}\!\!\!dx~
\partial_t[{\rm e}^{-\alpha\chi(x)}n^{\widehat{+}}(x)] \nonumber \\
&=&-i\frac{\lambda}{16\pi^2L}\int^\infty_{-\infty}\!\!\!dx~\partial_x
[{\rm e}^{\alpha\chi}(n^3\partial_xn^{{+}}-n^{{+}}\partial_xn^3)]
-i\frac{\sqrt{\lambda}}{4\pi}\sqrt{C}\int^\infty_{-\infty}\!\!\!dx~
\partial_x[{\rm e}^{-\alpha\chi}n^{{+}}]\nonumber\\&=&0\,,\nonumber\\
\label{partial_ttQ+_su2}
\partial_t\widetilde{Q}^{{-}}&=&-\frac{L}{2} \int^\infty_{-\infty}\!\!\!dx
~\partial_t[{\rm e}^{-\alpha\chi(x)}n^{\widehat{-}}(x)] \nonumber \\
&=&-i\frac{\lambda}{16\pi^2L}\int^\infty_{-\infty}\!\!\!dx~\partial_x
[{\rm e}^{\alpha\chi}(n^{{-}}\partial_xn^3-n^3\partial_xn^{{-}})]+i\frac{\sqrt{\lambda}}{4\pi}\sqrt{C}
\int^\infty_{-\infty}\!\!\!dx~\partial_x[{\rm e}^{-\alpha\chi}n^{{-}}]\nonumber\\&=&0\,.\nonumber
\label{partial_ttQ-_su2}
\end{eqnarray}
Thus $\widetilde{Q}^{{\pm}}$ are conserved. 

\medskip 

Note that all of the non-local charges are finite, 
while the local charge diverges.

\subsection{Null-like warped $SL(2)$ LLSM}

Finally, we consider the null-like warped $SL(2)$ LLSM. 

\medskip 

For this case, let us introduce the following boundary condition:
\begin{eqnarray}
n^-(x)\,,~~n^2(x) \rightarrow 0 \quad
\textrm{as} \quad x\rightarrow \pm \infty\,.
\label{condxxn}
\end{eqnarray}
In terms of the coordinates, this condition (\ref{condxxn}) is realized as the condition,
\begin{eqnarray}
\rho(x) \rightarrow \infty\,, \quad\psi(x)\rightarrow\frac{3}{2}\pi \quad
\textrm{as} \quad x\rightarrow \pm \infty\,,
\end{eqnarray}
where $\rho$ is supposed to diverge logarithmically. 
From the relation
\begin{eqnarray}
2n^+n^--\left(n^2\right)^2=1\,,
\end{eqnarray}
the condition (\ref{condxxn}) indicates that $n^+$ should diverge as $x\to\pm\infty$\,. 
\begin{eqnarray}
n^+(x)\rightarrow -\infty \qquad
\textrm{as} \qquad x\rightarrow \pm \infty\,.
\end{eqnarray}
Then the non-local charges $Q^+$ and $\widetilde{Q}^+$ are not finite 
and diverge. Hence a different component of the non-local charges 
diverges in comparison to the previous two cases. 
The difference comes from a pp-wave like limit and 
the divergence may be basically interpreted as an infinite length of the spatial 
direction of the string world-sheet.

\medskip

One can check that $Q^{+}$ and $Q^2$ are conserved as follows:   
\begin{eqnarray}
\partial_tQ^{+}&=&-\frac{L}{2} \int^\infty_{-\infty}\!\!\!dx~\partial_t[{\rm e}^{\sqrt{C}\alpha\chi(x)}n^{+}(x)] 
\nonumber \\
&=&\frac{\lambda}{16\pi^2L}\int^\infty_{-\infty}\!\!\!dx~
\partial_x[{\rm e}^{\sqrt{C}\alpha\chi}(n^2\partial_xn^+-n^+\partial_xn^2)]
+\frac{\sqrt{\lambda}}{4\pi}\sqrt{C}\int^\infty_{-\infty}\!\!\!dx~\partial_x[{\rm e}^{\sqrt{C}\alpha\chi}n^2]
\nonumber\\&=&0\,,\nonumber\\
\label{partialtQ+}
\partial_tQ^{2}&=&-\frac{L}{2} \int^\infty_{-\infty}\!\!\!dx~\partial_t[{\rm e}^{\sqrt{C}\alpha\chi(x)}n^{2}(x)]
\nonumber \\
&=&\frac{\lambda}{16\pi^2L}\int^\infty_{-\infty}\!\!\!dx~\partial_x[{\rm e}^{\sqrt{C}
\alpha\chi}(n^-\partial_xn^+-n^+\partial_xn^-)]+\frac{\sqrt{\lambda}}{4\pi}
\sqrt{C}\int^\infty_{-\infty}\!\!\!dx~\partial_x[{\rm e}^{\sqrt{C}\alpha\chi}n^-]
\nonumber\\&=&0\,. \nonumber
\label{partialtQ2}
\end{eqnarray}
Under the condition (\ref{condxxn}), all of the surface terms have vanished. 
Note that the differential terms are ensured to vanish due to the logarithmic behavior 
of $\rho$ around the spatial infinities.

\medskip 

The conservation laws of $\widetilde{Q}^+$ and $\widetilde{Q}^2$ are also shown as follows:
\begin{eqnarray}
\partial_t\widetilde{Q}^{+}&=&-\frac{L}{2} \int^\infty_{-\infty}\!\!\!dx~
\partial_t[{\rm e}^{-\sqrt{C}\alpha\chi(x)}n^{+}(x)] \nonumber \\
&=&\frac{\lambda}{16\pi^2L}\int^\infty_{-\infty}\!\!\!dx~\partial_x[{\rm e}^{-\sqrt{C}
\alpha\chi}(n^2\partial_xn^+-n^+\partial_xn^2)]-\frac{\sqrt{\lambda}}{4\pi}
\sqrt{C}\int^\infty_{-\infty}\!\!\!dx~\partial_x[{\rm e}^{-\sqrt{C}\alpha\chi}n^2]
\nonumber\\&=&0\,,\nonumber\\
\partial_t\widetilde{Q}^{2}&=&-\frac{L}{2} \int^\infty_{-\infty}\!\!\!dx~
\partial_t[{\rm e}^{-\sqrt{C}\alpha\chi(x)}n^{2}(x)] \nonumber \\
&=&\frac{\lambda}{16\pi^2L}\int^\infty_{-\infty}\!\!\!dx~\partial_x[{\rm e}^{-\sqrt{C}\alpha\chi}
(n^-\partial_xn^+-n^+\partial_xn^-)]-\frac{\sqrt{\lambda}}{4\pi}\sqrt{C}\int^\infty_{-\infty}\!\!\!dx~
\partial_x[{\rm e}^{-\sqrt{C}\alpha\chi}n^-]\nonumber\\&=&0\,.\nonumber
\end{eqnarray}
As a result, $\widetilde{Q}^+$ and $\widetilde{Q}^2$ are conserved 
under the condition (\ref{condxxn}).

\medskip 

Note that the non-local charges $Q^2$ and $\widetilde{Q}^2$ as well as 
the local charge $Q^-$ are finite. However, the non-local charges 
$Q^+$ and $\widetilde{Q}^+$ diverge.

\section{Null-like warped LLSM from time-like LLSM }

We derive here the null-like warped $SL(2)$ LLSM from the time-like warped $SL(2)$ LLSM 
by taking an appropriate scaling-limit (a pp-wave like limit). 
The classical equations of motion are first reproduced 
and then the Lax pair is also derived. 

\subsection{Equations of motion}
 
The equations of motion of the null-like warped $SL(2)$ LLSM are reproduced 
from those of the time-like warped $SL(2)$ LLSM by taking a scaling limit. 

\medskip 

Recall the equations of motion for the time-like warped $SL(2)$ LLSM,
\begin{eqnarray}
&&\partial_t n^0=\frac{\lambda}{8\pi^2L^2}\left(n^1\partial_x^2 n^2-n^2\partial_x^2 n^1\right)\,, 
\nonumber \\
&&\partial_t n^1=\frac{\lambda}{8\pi^2L^2}\left(n^0\partial_x^2 n^2-n^2\partial_x^2 n^0\right)-2\widetilde{C}n^0n^2\,, 
\nonumber \\
&&\partial_t n^2=\frac{\lambda}{8\pi^2L^2}\left(n^1\partial_x^2 n^0-n^0\partial_x^2 n^1\right)+2\widetilde{C}n^0n^1\,.
\end{eqnarray}
Here we have used $\widetilde{C}$ as a deformation parameter.   
It is convenient to introduce $n^\pm$ as 
\begin{eqnarray}
n^\pm=\frac{n^0\pm n^1}{\sqrt{2}}\,. 
\end{eqnarray}
Then let us rewrite the equations of motion in terms of $n^\pm$ 
with the following rescaling: 
\begin{eqnarray}
n^-\to\sqrt{\frac{2C}{\widetilde{C}}}\,n^-\,, \qquad
n^+\to\sqrt{\frac{\widetilde{C}}{2C}}\,n^+\,.
\end{eqnarray}
Finally, by taking the $\widetilde{C}\to 0$ limit with $C$ fixed, 
the equations of motion are evaluated as 
\begin{eqnarray}
&&\partial_t n^-=-\frac{\lambda}{8\pi^2L^2}\left(n^-\partial_x^2 n^2-n^2\partial_x^2 n^-\right)\,, \nonumber \\
&&\partial_t n^2=-\frac{\lambda}{8\pi^2L^2}\left(n^-\partial_x^2 n^+-n^+\partial_x^2 n^-\right)-2C\left(n^-\right)^2\,, \nonumber \\
&&\partial_t n^+=-\frac{\lambda}{8\pi^2L^2}\left(n^2\partial_x^2 n^+-n^+\partial_x^2 n^2\right)-2Cn^-n^2\,. 
\end{eqnarray}
These are equivalent to the equations of motion for the null-like warped LLSM in (\ref{eomXXN}).

\subsection{A Lax pair of null-like warped $SL(2)$ LLSM}

Let us derive the Lax pair of the null-like warped $SL(2)$ LLSM. 

\medskip 

We start from the Lax pair of the time-like warped $SL(2)$ LLSM. 
\begin{eqnarray}
U(t,x;z)&=&\dfrac{i\widetilde{\alpha}}{\sinh z} \left[-\cosh z\, n^0T^0 + n^1T^1+ n^2T^2 \right] \,, \label{C5}\\
V(t,x;z)&=&\dfrac{i\widetilde{\beta}}{\sinh z} \left[-\cosh z\left(n^1\partial_x n^2-n^2\partial_x n^1\right)T^0 +\left(n^0\partial_x n^2-n^2\partial_x n^0\right) T^1 \right. \nonumber \\
&&\hspace{1.5cm}\left.+\left(n^1\partial_x n^0-n^0\partial_x n^1\right)T^2 \right] \nonumber \\
&&+\dfrac{\widetilde{\alpha}\widetilde{\beta}}{\sinh^2 z} \left[- n^0T^0 + \cosh z\,n^1T^1+\cosh z\,n^2T^2 \right]\,. \notag 
\end{eqnarray}
Here $\widetilde{\alpha}$ and $\widetilde{\beta}$ are defined as 
\begin{eqnarray}
\widetilde{\alpha} \equiv \frac{4\pi L}{\sqrt{\lambda}}\sqrt{\widetilde{C}}\,, \qquad
\widetilde{\beta} \equiv \frac{\sqrt{\lambda}}{2\pi L}\sqrt{\widetilde{C}}\,,  
\end{eqnarray}
and $\widetilde{C}$ is a deformation parameter. 

\medskip 

Let us rewrite the Lax pair (\ref{C5}) in terms of $T^\pm$ and  
then rescale $z$\,, $n^\pm$ and $T^\pm$ like  
\begin{eqnarray}
&& z\to i\sqrt{\widetilde{C}}\,z\,, \quad 
n^- \to \sqrt{\frac{2C}{\widetilde{C}}}\,n^-\,, \quad 
n^+ \to \sqrt{\frac{\widetilde{C}}{2C}}\,n^+\,, \nonumber \\
&& T^- \to \sqrt{\frac{2C}{\widetilde{C}}}\,T^-\,, \quad 
T^+ \to \sqrt{\frac{\widetilde{C}}{2C}}\,T^+\,. 
\end{eqnarray} 
Finally, by taking the $\widetilde{C} \to 0$ limit, the following Lax pair is obtained,  
\begin{eqnarray}
U(t,x;z)&=&\dfrac{\alpha}{z}
\left(-\left[n^+-\frac{Cz^2}{2}n^-\right]T^-
+n^2T^2-n^-T^+\right) \,, \nonumber \\
V(t,x;z)&=&\dfrac{\beta}{z}
\Biggl(-\left[\left(n^2\partial_x n^+-n^+\partial_x n^2\right)
-\frac{Cz^2}{2}\left(n^-\partial_x n^2-n^2\partial_x n^-\right)\right]T^- \Biggr. \nonumber \\
&&\hspace{1cm}+\Biggl. \left(n^-\partial_xn^+-n^+\partial_xn^-\right)T^2
-\left(n^-\partial_x n^2-n^2\partial_x n^-\right)T^+\Biggr) \nonumber \\
&&+\dfrac{\alpha\beta}{z^2} 
\left(-\left[n^++\frac{Cz^2}{2}n^-\right]T^-+n^2T^2-n^-T^+\right)\,. \label{C8}
\end{eqnarray}
Here $\alpha$ and $\beta$ have newly been introduced as 
\begin{eqnarray}
\alpha=\frac{4\pi L}{\sqrt{\lambda}}\,, \qquad
\beta=-\frac{\sqrt{\lambda}}{2\pi L}\,. 
\end{eqnarray}
From the Lax pair (\ref{C8}), 
one can reproduce the equations of motion 
for the null-like warped $SL(2)$ LLSM obtained in \cite{KameYoshi}.

\section{Undoing Jordanian twists}

We present here non-local gauge transformations for the Lax pair of the null-like warped $SL(2)$ LLSM. 
The gauge transformations may be interpreted as undoing Jordanian twists. 

\medskip 

Let us start from the following Lax pair,  
\begin{eqnarray}
U(t,x;z)&=&\dfrac{\alpha}{z}
\left(-\left[n^+-\frac{Cz^2}{2}n^-\right]T^-
+n^2T^2-n^-T^+\right) \,, \nonumber \\
V(t,x;z)&=&\dfrac{\beta}{z}
\Biggl(-\left[\left(n^2\partial_x n^+-n^+\partial_x n^2\right)
-\frac{Cz^2}{2}\left(n^-\partial_x n^2-n^2\partial_x n^-\right)\right]T^- \Biggr. \nonumber \\
&&\hspace{1cm}+\Biggl. \left(n^-\partial_xn^+-n^+\partial_xn^-\right)T^2
-\left(n^-\partial_x n^2-n^2\partial_x n^-\right)T^+\Biggr) \nonumber \\
&&+\dfrac{\alpha\beta}{z^2} 
\left(-\left[n^++\frac{Cz^2}{2}n^-\right]T^-+n^2T^2-n^-T^+\right)\,. \label{C1}
\end{eqnarray}
Note that this Lax pair has two poles at $z=0$ and $z=\infty$\,. 

\medskip 

The isomorphisms of $sl(2)$ algebra, 
\begin{eqnarray}
T^-\to T^-\,, \quad
T^2\to T^2 \mp \sqrt{C}zT^-\,, \quad
T^+\to T^+ \mp \sqrt{C}zT^2 +\frac{Cz^2}{2}T^-\,, 
\end{eqnarray}
transforms the Lax pair (\ref{C1}) into 
\begin{eqnarray}
U^{(\pm)}(t,x;z) &=&\dfrac{\alpha}{z}
\left[-\left(n^+\pm\sqrt{C}zn^2\right)T^-
+\left(n^2\pm\sqrt{C}zn^-\right)T^2-n^-T^+\right] \,, \\
V^{(\pm)}(t,x;z) &=&\dfrac{\beta}{z}
\biggl(-\left[\left(n^2\partial_x n^+-n^+\partial_x n^2\right)
\pm\sqrt{C}z\left(n^-\partial_x n^+-n^+\partial_x n^-\right)\right]T^- \Biggr. \nonumber \\
&&\hspace{0.8cm}+\left[\left(n^-\partial_xn^+-n^+\partial_xn^-\right)
\pm\sqrt{C}z\left(n^-\partial_x n^2-n^2\partial_x n^-\right)\right]T^2 \nonumber \\
&&\hspace{0.8cm}\Biggl.-\left(n^-\partial_x n^2-n^2\partial_x n^-\right)T^+ \biggr) \nonumber \\
&&+\dfrac{\alpha\beta}{z^2} 
\left(-\left[n^+\pm\sqrt{C}zn^2+Cz^2n^-\right]T^-+\left(n^2\pm\sqrt{C}zn^-\right)T^2-n^-T^+
\right)\,. \nonumber
\end{eqnarray}
The resulting Lax pairs $(U^{(\pm)}(z),V^{(\pm)}(z))$ do not diverge at $z=\infty$ any more.    
Then the asymptotic expressions are given by  
\begin{eqnarray}
U^{(\pm)}(t,x;z=\infty)&=&\pm\sqrt{C}\alpha\left(-n^2T^-+n^-T^2\right) \,, \label{C4} \\
V^{(\pm)}(t,x;z=\infty)&=&\pm\sqrt{C}\beta
\left[-\left(n^-\partial_x n^+-n^+\partial_x n^-\right)T^- 
\right. \nonumber \\ &&\hspace{1.5cm}\left.
+\left(n^-\partial_x n^2-n^2\partial_x n^-\right)T^2\right]
-C\alpha\beta n^-T^-\,. \nonumber
\end{eqnarray}
These are useful to construct non-local gauge transformations. 

\medskip 

With the help of (\ref{C4}), non-local functions can be defined as 
\begin{eqnarray}
{\mathcal F}^{(\pm)}(t,x) \equiv {\rm P}\exp\left[\int^x_{-\infty}\!\!\!dy~U^{(\pm)}(t,y;z=\infty)\right]K^{(\pm)}\,. 
\label{sol} 
\end{eqnarray}
These functions are obtained as the solutions of the following differential equations: 
\begin{eqnarray}
\partial_t{\mathcal F}^{(\pm)}=V^{(\pm)}(z=\infty){\mathcal F}^{(\pm)}\,, \qquad
\partial_x{\mathcal F}^{(\pm)}=U^{(\pm)}(z=\infty){\mathcal F}^{(\pm)}\,. 
\end{eqnarray}
When solving the differential equations, $K^{(\pm)}$ 
are introduced as integral constant matrices. 
It is convenient to take $K^{(\pm)}$ as follows: 
\begin{eqnarray}
K^{(\pm)}= \exp\left[\pm \dfrac{2\sqrt{C}\alpha}{L} Q^- T^2\right]\,. 
\end{eqnarray}
These choices are basically fixed in \cite{Jordanian-KMY} by borrowing the knowledge 
of quantum Jordanian twists. 
Then the explicit expressions of ${\mathcal F}^{(\pm)}$ are given by, respectively,   
\begin{eqnarray}
&&{\mathcal F}^{(+)}=\exp\left[\sqrt{C}\alpha\chi(x) T^2\right]
\exp\left[-\sqrt{C}\alpha\left(\xi(x)-\frac{1}{L}Q^2\right)T^-\right]
\exp\left[\frac{\sqrt{C}\alpha}{L}Q^-T^2\right]\,, \\
&&{\mathcal F}^{(-)}=\exp\left[-\sqrt{C}\alpha\chi(x) T^2\right]
\exp\left[\sqrt{C}\alpha\left(\widetilde{\xi}(x)-\frac{1}{L}\widetilde{Q}^2\right)T^-\right]
\exp\left[-\frac{\sqrt{C}\alpha}{L}Q^-T^2\right]\,. \nonumber
\end{eqnarray}
where $\xi(x)$ and $\widetilde{\xi}(x)$ are non-local fields defined as 
\begin{eqnarray}
&&\xi(x) \equiv  \frac{1}{2}\int^\infty_{-\infty}\!\!\!dy~\epsilon(x-y){\rm e}^{\sqrt{C}\alpha\chi(x)}n^{2}(y)\,,\\
&&\widetilde{\xi}(x) \equiv \frac{1}{2}\int^\infty_{-\infty}\!\!\!dy~\epsilon(x-y){\rm e}^{-\sqrt{C}\alpha\chi(x)}n^{2}(y)\,.  \notag
\end{eqnarray}

\medskip 

The next is to see the transformation laws of the Lax pairs $(U^{(\pm)}(z),V^{(\pm)}(z))$ under 
gauge transformations generated by ${\mathcal F}^{(\pm)}$\,, 
\begin{eqnarray}
{\mathcal U}^{(\pm)}(z)&=&\left({\mathcal F}^{(\pm)}\right)^{-1}U(z){\mathcal F}^{(\pm)}
-\left({\mathcal F}^{(\pm)}\right)^{-1}\partial_x\,{\mathcal F}^{(\pm)}\,, \\
{\mathcal V}^{(\pm)}(z)&=&\left({\mathcal F}^{(\pm)}\right)^{-1}V(z){\mathcal F}^{(\pm)}
-\left({\mathcal F}^{(\pm)}\right)^{-1}\partial_t\,{\mathcal F}^{(\pm)}\,. \nonumber
\end{eqnarray}
The resulting Lax pairs $({\mathcal U}^{(\pm)}(z),{\mathcal V}^{(\pm)}(z))$ are explicitly given by 
\begin{eqnarray}
{\mathcal U}^{(+)}(z)&=&\dfrac{\alpha}{z}\left[-{\mathcal N}^-T^++{\mathcal N}^2T^2-{\mathcal N}^+T^-\right]\,, \\
{\mathcal V}^{(+)}(z)&=&\dfrac{\beta}{z}
\Bigl[-\left({\mathcal N}^-\partial_x {\mathcal N}^2-{\mathcal N}^2\partial_x {\mathcal N}^-\right)T^+
+\left({\mathcal N}^-\partial_x{\mathcal N}^+-{\mathcal N}^+\partial_x{\mathcal N}^-\right)T^2
 \nonumber \\
&&\hspace{0.8cm}\Bigl.-\left({\mathcal N}^2\partial_x {\mathcal N}^+-{\mathcal N}^+\partial_x {\mathcal N}^2\right)T^-\Bigr] 
+\dfrac{\alpha\beta}{z^2} 
\Bigl[-{\mathcal N}^-T^++{\mathcal N}^2T^2-{\mathcal N}^+T^-\Bigr]\,, \nonumber \\
{\mathcal U}^{(-)}(z)&=&\dfrac{\alpha}{z}\left[-\widetilde{\mathcal N}^-T^++\widetilde{\mathcal N}^2T^2-\widetilde{\mathcal N}^+T^-\right]\,, \nonumber \\
{\mathcal V}^{(-)}(z)&=&\dfrac{\beta}{z}
\Bigl[-\left(\widetilde{\mathcal N}^-\partial_x \widetilde{\mathcal N}^2-\widetilde{\mathcal N}^2\partial_x \widetilde{\mathcal N}^-\right)T^+
+\left(\widetilde{\mathcal N}^-\partial_x\widetilde{\mathcal N}^+-\widetilde{\mathcal N}^+\partial_x\widetilde{\mathcal N}^-\right)T^2
 \nonumber \\
&&\hspace{0.8cm}\Bigl.-\left(\widetilde{\mathcal N}^2\partial_x \widetilde{\mathcal N}^+-
\widetilde{\mathcal N}^+\partial_x \widetilde{\mathcal N}^2\right)T^-\Bigr] 
+\dfrac{\alpha\beta}{z^2} 
\Bigl\{-\widetilde{\mathcal N}^-T^++\widetilde{\mathcal N}^2T^2-\widetilde{\mathcal N}^+T^-\Bigr\}\,, \nonumber
\end{eqnarray}
where the components of non-local vectors ${\mathcal N}^a$ and $\widetilde{\mathcal N}^a$ are given by 
\begin{eqnarray}
&&{\mathcal N}^-
=-\frac{1}{\sqrt{C}\alpha}\textrm{e}^{-\frac{\sqrt{C}\alpha}{L}Q^-}\partial_x\textrm{e}^{-\sqrt{C}\alpha\chi(x)}\,, \quad
{\mathcal N}^2=\partial_x\left[\left(\xi-\frac{1}{L}Q^2\right)\textrm{e}^{-\sqrt{C}\alpha\chi(x)}\right]\,, \\
&&{\mathcal N}^+=\textrm{e}^{\frac{\sqrt{C}\alpha}{L}Q^-}\left(\textrm{e}^{\sqrt{C}\alpha\chi(x)}n^+(x)
-\frac{\sqrt{C}\alpha}{2}\partial_x\left[\left(\xi-\frac{1}{L}Q^2\right)^2
\textrm{e}^{-\sqrt{C}\alpha\chi(x)}\right]\right)\,. \nonumber \\
&&\widetilde{\mathcal N}^-=\frac{1}{\sqrt{C}\alpha}\textrm{e}^{\frac{\sqrt{C}\alpha}{L}Q^-}
\partial_x\textrm{e}^{\sqrt{C}\alpha\chi(x)}\,, \quad
\widetilde{\mathcal N}^2=\partial_x\left[\left(\widetilde{\xi}-\frac{1}{L}\widetilde{Q}^2\right)
\textrm{e}^{\sqrt{C}\alpha\chi(x)}\right]\,, \nonumber \\
&&\widetilde{\mathcal N}^+=\textrm{e}^{-\frac{\sqrt{C}\alpha}{L}Q^-}
\left(\textrm{e}^{-\sqrt{C}\alpha\chi(x)}n^+(x) + \frac{\sqrt{C}\alpha}{2}\partial_x
\left[\left(\widetilde{\xi}-\frac{1}{L}\widetilde{Q}^2\right)^2
\textrm{e}^{\sqrt{C}\alpha\chi(x)}\right]\right)\,. \nonumber
\end{eqnarray}
They satisfy the following relations,
\begin{eqnarray}
2{\mathcal N}^-{\mathcal N}^+ -\left({\mathcal N}^2\right)^2=1\,,\qquad 2\widetilde{\mathcal N}^-
\widetilde{\mathcal N}^+ -\bigl(\widetilde{\mathcal N}^2\bigr)^2=1\,.
\end{eqnarray}

\medskip 

The monodromy matrices constructed from the Lax pairs $(\mathcal{U}^{(\pm)}(z),\mathcal{V}^{(\pm)}(z))$ 
lead to a couple of Yangians, as shown in the body of this manuscript. 
This result indicates that the non-local gauge transformations are nothing but undoing Jordanian twists.

\end{document}